\documentclass[a4paper,12pt]{article}

\usepackage{amssymb}
\usepackage{amsmath}
\usepackage{graphicx}
\usepackage[round,comma]{natbib}
\usepackage[colorlinks]{hyperref}
\hypersetup{colorlinks,breaklinks=true,citecolor=blue,filecolor=blue,linkcolor=blue,urlcolor=blue,pdfauthor={C.~Guervilly}}

\newcommand{\vel}{\mathbf{u}}
\newcommand{\B}{\mathbf{B}}
\newcommand{\cur}{\mathbf{j}}
\newcommand{\pdt}[1]{\frac{\partial #1}{\partial t}}
\newcommand{\n}{\nabla}

\newcommand{\pl}{\left(}
\newcommand{\pr}{\right)}

\title{Numerical Simulations of Dynamos Generated in Spherical Couette Flows}
\author{C\'eline Guervilly \& Philippe Cardin}
\date{October 20, 2010}
\begin{document}

\maketitle
\begin{abstract}
We numerically investigate the efficiency of a spherical Couette flow at generating a self-sustained magnetic field. No dynamo action occurs for axisymmetric flow while we always found a dynamo when non-axisymmetric hydrodynamical instabilities are excited. Without rotation of the outer sphere, typical critical magnetic Reynolds numbers $Rm_c$ are of
the order of a few thousands. They increase as the mechanical forcing imposed by the inner core on the flow increases (Reynolds number $Re$). Namely, no dynamo is found if the magnetic Prandtl number $Pm=Rm/Re$ is less than a critical value $Pm_c\sim 1$. Oscillating quadrupolar dynamos are present in the vicinity of the dynamo onset. 
Saturated magnetic fields obtained in supercritical regimes (either $Re>2 Re_c$ or
$Pm>2Pm_c$) correspond to the equipartition between magnetic and kinetic energies. A global rotation of the system (Ekman numbers $E=10^{-3}, 10^{-4}$ ) yields to a slight decrease (factor 2) of the critical magnetic Prandtl number, but we find a peculiar regime where dynamo action may be obtained for relatively low magnetic Reynolds numbers
($Rm_c\sim 300$). In this dynamical regime (Rossby number $Ro\sim -1$, spheres in opposite direction) at a moderate Ekman number ($E=10^{-3}$), a enhanced shear layer around the inner core might explain the decrease of the dynamo threshold. For lower $E$ ($E=10^{-4}$) this internal shear layer becomes unstable, leading to small scales
fluctuations, and the favorable dynamo regime is lost. We also model the effect of ferromagnetic boundary conditions. Their presence have only a small impact on the dynamo onset but clearly enhance the saturated magnetic field in the ferromagnetic parts. Implications for experimental studies are discussed.

\end{abstract}

\section{Introduction}
Naturally occuring dynamos are driven by the motions of metallic fluids of very low magnetic Prandtl number $Pm$, the ratio of kinematic viscosity to magnetic diffusivity ($Pm\le 10^{-5}$). The liquid metals used in laboratory dynamo experiments have also this property. Accordingly, many groups in the world have devised experiments using liquid metals to model and study naturally occuring dynamos. The first generation of dynamo experiments, performed ten years ago, used constrained flows as close as possible to well known kinematic dynamos. In two experiments using liquid sodium, one in Riga, Latvia \citep{Gai01} and one in Karlsruhe, Germany \citep{Sti01}, the dynamo effect has indeed been observed. More recently, second-generation experiments have been performed in order to study the amplification of the magnetic field by less constrained flows. The flow is highly turbulent but the mean flow is expected to yield dynamo action \citep{Pef00,Bou02,For02}. No self-sustained magnetic fields were observed in the parameter regime where kinematic numerical computations predicted dynamos. It seems that hydrodynamic turbulence inhibits dynamo action in these experiments, at least close to the predicted dynamo onset. In 2007 however due to the use of ferromagnetic impellers, a self-sustained magnetic field has been observed in the Von K\'arm\'an Sodium (VKS) experiment in Cadarache \citep{Mon07}. Whereas in natural systems such as the Earth's core the magnetic field is strong and, as a result, modifies largely the velocity field by Lorentz forces, in all these experiments the induced magnetic field is weak and do not strongly act back on the flow. Consideration of key ingredients for the flow dynamics (strong magnetic field, overall rotation of the system) in natural systems has motivated the construction of the third generation of experiments. Spherical Couette flows (produced between two concentric spheres in differential rotation) are geometrically relevant for the Earth's core and may incorporate global rotation (i.e. rotation of the outer sphere). Tests are currently underway on experimental spherical Couette setups \citep{Sis04,Nat06}. Fully three-dimensional numerical simulations can inform such experiments. Computing dynamos at low $Pm$ is very challenging as the magnetic Reynolds number $Rm$ required for dynamo action is at least $10-100$, and hence the Reynolds number $Re=Rm/Pm$ should be $10^6-10^7$ with $Pm\sim 10^{-5}$ as for liquid sodium. High $Re$ flows are turbulent and so computationally demanding. To overcome this difficulty $Pm$ is taken to be order unity in dynamo simulations.
\newline \indent 
Spherical Couette flow may also be relevant to giant planets: bands of strong zonal winds are visible at their surfaces, and if deep rooted \citep{Hei05} they can exert a forcing on the outer boundary of the core. 
\newline \indent 
Non-conducting Couette flows have been numerically investigated \citep{Hol06,Hol04} but surprisingly no study has been carried out for conducting flow generating an induced magnetic field without the presence of an imposed magnetic field \citep{Hol07}.  Magnetized Taylor-Couette flows (produced between two coaxial cylinders in differential rotation) have been widely investigated, mostly to model magneto-rotational instability in Keplerian disks \citep[e.g.][]{Hol05}. The existence of self-generation of magnetic field in Taylor-Couette flow has been numerically demonstrated \citep{Lau00,Wil02}. 
\newline \indent
In this work we test the ability of a spherical Couette flow to generate self-sustained magnetic fields. We investigate a large range of parameter values in order to determine the most favourable regime for dynamo action. Before investigating the influence of global rotation on the flow dynamics and dynamo action we first consider the case with the outer sphere held at rest. In the second part of this paper, for experimental interests, we address the problem of the impact of high magnetic permeability boundary conditions on the dynamo threshold.

\section{Governing equations}
A solid inner sphere of radius $r_i$ is contained in a sphere of radius $r_o$. The two spheres rotate with the same axis of rotation $\mathbf{e_z}$. The solid inner core rotates faster or slower than the outer sphere at the rotation rate $\Omega+\Delta \Omega$ where $\Omega$ is the rotation rate of the outer sphere. The aspect ratio $\chi =r_i/r_o$ is $0.35$, as in the Earth's core. The spherical shell is filled with an incompressible fluid of viscosity $\nu$. This fluid can be non-conducting to model a fluid as water or conducting as liquid metal with an electrical conductivity $\sigma=(\mu_0 \eta)^{-1}$, with $\mu_0$ the vaccum magnetic permeability and $\eta$ the magnetic diffusivity. The density of the fluid $\rho$ is uniform; the temperature is assumed to be uniform and constant throughout the simulation. The fluid is driven by differential rotation between the two spheres. Although in an experiment the inner sphere would be supported by one or two shafts, these are not modelled in our numerical simulations.
\newline \indent
In the first part of this paper (sections~\ref{sec:E0} and~\ref{sec:E}) the inner core is assumed to have the same electric and magnetic properties as the fluid, i.e. the same electrical conductivity and magnetic permeability $\mu_0$. The outer sphere is an electric insulator with a magnetic permeability $\mu_0$. In the second part (section~\ref{sec:ferro}) the magnetic permeabilities of the inner core and the outer sphere are varied. A seed magnetic field is imposed at initial time, and freely evolved during the time integration. The velocity boundary conditions are no slip. We work in the frame rotating with the outer boundary.
\newline \indent
The velocity $\vel$ is scaled by $r_i\Delta\Omega$, the inner core azimuthal velocity at the equator in the rotating frame. The length scale is the radius of the outer sphere $r_o$. $\B$ is scaled by $\sqrt{\rho \mu_0 r_i r_o (\Omega+\Delta \Omega) \Delta\Omega}$. The time evolution of the divergence-free velocity field $\vel$ and the divergence-free magnetic field $\B$ are given by the Navier-Stokes and the magnetic induction equations respectively:
\begin{equation}
Re \pdt{\vel}+ Re \pl \vel\cdot\boldsymbol{\n} \pr \vel + 2 \frac{1}{E} \mathbf{e_z}\times \vel 
= -\boldsymbol{\n}\Pi +\boldsymbol{\n^2}\vel+ \pl \frac{1}{E} + \frac{Re}{\chi} \pr \pl \boldsymbol{\n} \times \B \pr \times \B ,
\label{eq:NS}
\end{equation}
\begin{equation}
\pdt \B = \boldsymbol{\n} \times \pl \vel \times \B \pr + \frac{1}{Re Pm} \boldsymbol{\n^2}\B ,
\label{eq:ind}
\end{equation}
where $\Pi$ is the dimensionless pressure.
\newline \indent  
The Reynolds number $Re=r_o r_i\Delta \Omega / \nu$ parametrizes the mechanical forcing exerted on the system by controling the rotation rate of the inner core. The magnetic Prandtl number $Pm=\nu / \eta$ influences the magnetic Reynolds number $Rm=Re Pm$ and hence affects the rate of magnetic induction. The Ekman number $E=\nu / (\Omega r_o^2)$ is an inverse measure of the overall rotation rate $\Omega$. We recall that the aspect ratio $\chi=r_i/r_o$ is $0.35$.
\newline \indent
The numerical implementation of the above equations uses the \texttt{PARODY} code, derived from \citet{Dor97}, Aubert, Cardin, Dormy in the dynamo benchmark \citep{Chr01}, and parallelized and optimized by J. Aubert and E. Dormy. The velocity and magnetic fields are decomposed into poloidal and toroidal scalars and expanded in spherical harmonic functions in the angular coordinates. A finite difference scheme is used on an irregular radial grid (thinner near the boundaries to resolve the boundary layers). A Crank-Nicolson scheme is implemented for the time integration of the diffusion terms and an Adams-Bashforth procedure is used for the other terms. All results presented in this paper arise from fully three-dimensional and non-linear computations. Typical resolutions are $150-300$ radial grid points and $64-200$ spherical harmonics degrees and orders. 

\section{Outer sphere at rest}
\label{sec:E0}
Without global rotation no Coriolis force acts on the system. We recall that the inner core has the same electric and magnetic properties as the fluid and the outer boundary is insulating. We first present the results obtained with a non-conducting fluid ($Pm=0$) (section~\ref{sec:hydroE0}) in order to determine the flow dynamics generating the dynamo mechanism when the fluid is conducting (section~\ref{sec:magE0}).

\subsection{Hydrodynamics}
\label{sec:hydroE0}
\begin{figure}
\begin{center}
	\resizebox*{12cm}{!}{\includegraphics{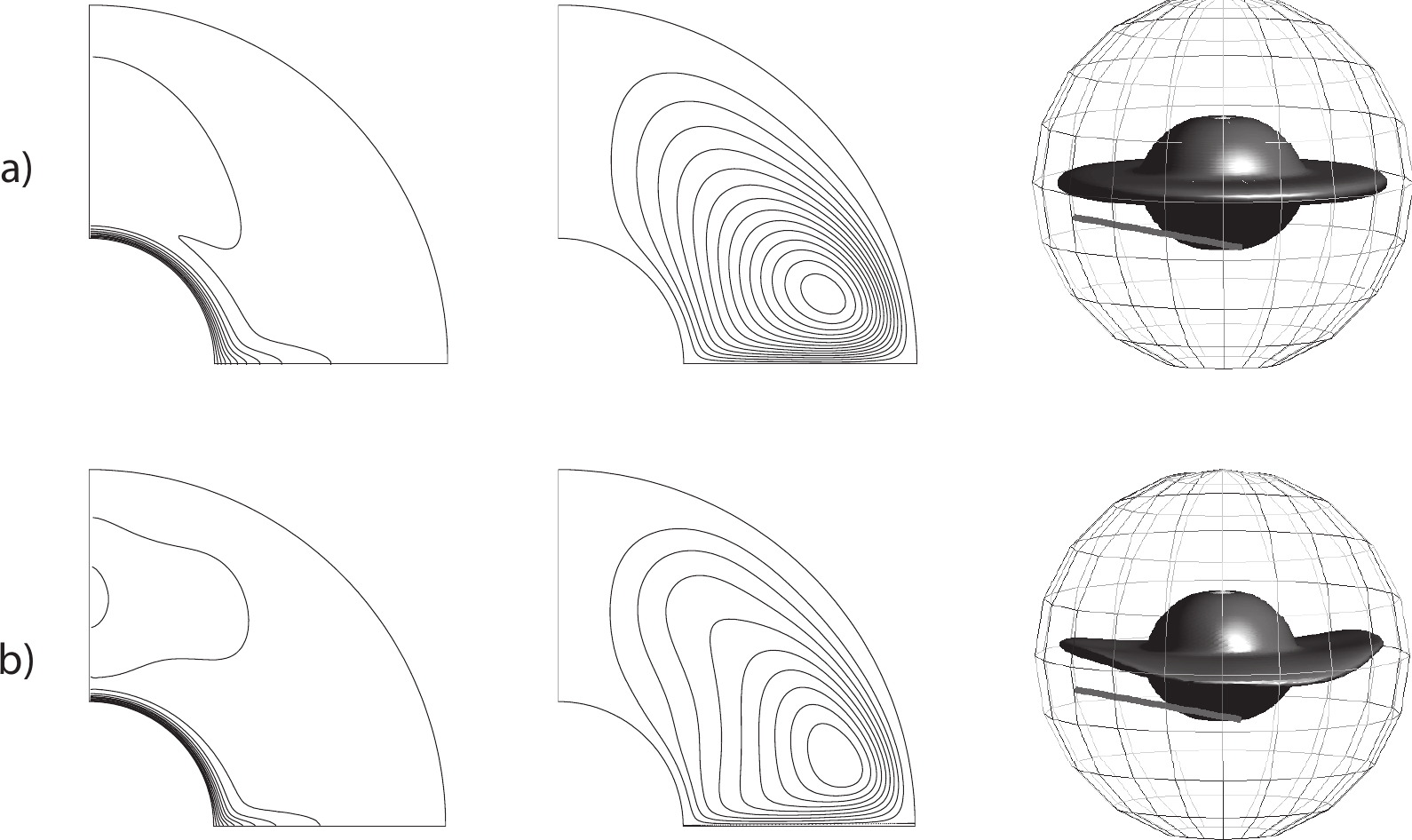}}
	\caption{Snapshots of contours of the angular velocity (zonally averaged) with a contour interval of $0.1$ (left), streamlines of the meridional circulation with a contour interval of $0.001$ (middle) plotted in a quadrant of the meridional plane and isolevel of the kinetic energy correponding to $20\%$ of the maximal value (right) for (a) a subcritical regime $Re=1000$ and (b) a supercritical regime $Re=2000$. For $Re=2000$ the most unstable mode is $m=2$. The localization of the beam where the velocity is measured in the DTO experiment \citep{Gag09} is drawn in gray thick line on the 3D-plots.}
	\label{fig:hat}
\end{center}
\end{figure}

Following earlier work by \citet{Dum91}, \citet{Hol06} investigated hydrodynamical spherical Couette flow for various aspect ratios. They found that the basic axisymmetric flow consists of a radial centrifugal jet concentrated in the equatorial plane, carrying fluid from the inner sphere to the outer one, and an azimuthal component of same order of magnitude for the aspect ratio we are interested in. When increasing the Reynolds number above a critical value ($Re_c$) instabilites develop, which are interpreted as series of longitudinal waves deforming the jet alternatively above and below the equatorial plane. These results are in good quantitative agreement with experimental studies \citep{Egb95}. For an aspect ratio equal to $r_i/r_o=0.35$, \citet{Hol06} found that the critical Reynolds number $Re_c$ (according to our definition of $Re$) is around $1100$ and the most unstable azimuthal wavenumber is $3$ at the instability onset. In the supercritical regime the most unstable mode decreases from $3$ to $2$.
\newline \indent
Figure~\ref{fig:hat}(a) shows the angular velocity contours, the meridional circulation and an isolevel of the kinetic energy for a subcritical regime ($Re<Re_c$). For $Re \le 1100$ the fluid flows from the pole to the equator close to the inner sphere and is centrifuged toward the outer sphere in the equatorial plane, forming an axisymmetric zonal jet. The recirculation takes place near the outer sphere from the equator to the pole forming a cell in each hemisphere. Moreover due to angular momentum transport along the meridional circulation streamlines a sizeable angular velocity is present in the polar regions. When the boundary forcing is increased ($Re\ge 1200$), as the basic axisymmetric flow evolves, a non-axisymmetric instability breaking the equatorial symmetry occurs as an azimuthal oscillation of the equatorial jet (figure~\ref{fig:hat}(b)). Note that the phase velocity of the oscillation does not undergo a phase shift with the radius as it can be seen in the three-dimensional plot of figure \ref{fig:hat}(b) and in \citet{Hol06} (no spiralling of the oscillating structure) whereas the angular velocity depends on the radius. Therefore the oscillations are interpretated as a wave propagating in the azimuthal direction, in opposition to a passive transport by the azimuthal flow. Nevertheless the issue of the mechanism triggering the instability remains unsolved. In the large gap study of \citet{Hol06}, a new kind of instability was found, leaving the role played by outer boundary open.

\begin{figure}
\begin{center}
	\resizebox*{15cm}{!}{\includegraphics{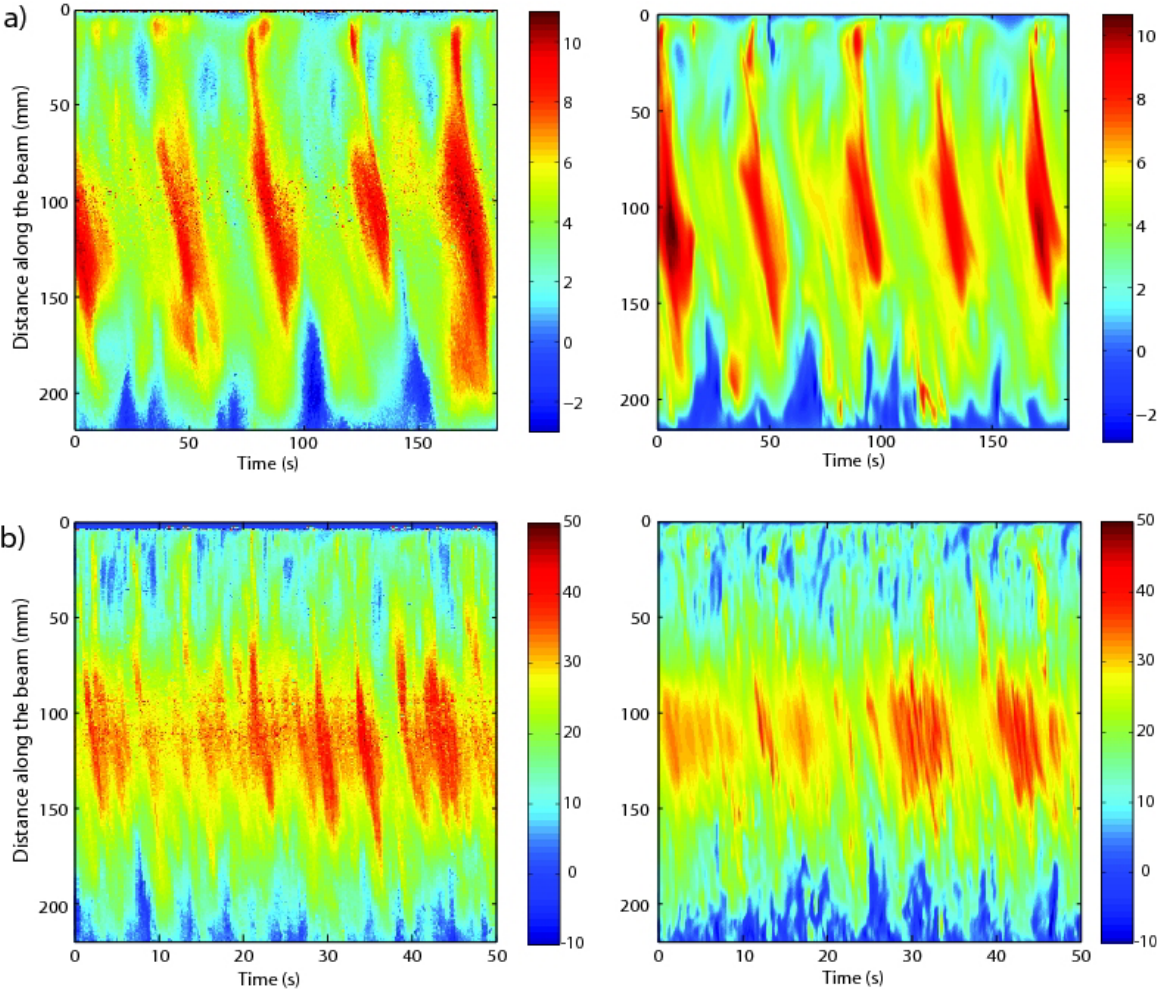}}
	\caption{Spatio-temporal diagrams of the velocity measured along a beam located in a plane parallel to the equatorial plane at the colatitude $103.9^\circ$ (see 3D-plots in figure \ref{fig:hat}) by ultrasonic Doppler velocimetry in the DTO experiment \citep{Gag09} (left) and computed in the numerical simulation (right) for (a) $Re=7200$ and (b) $Re=31678$.  The unity of the velocity is mm/s.}
	\label{fig:DTO}
\end{center}
\end{figure}

\indent 
Simulations at high Reynolds number have been performed and compared with experimental results obtained in a similar geometry by the Grenoble group \citep{Gag09}. The experiment (called DTO) consists of a $4.4$ cm-radius inner core surrounded by a $12$ cm-radius sphere in plexiglas. The working fluid is water with pollen particles in suspension. Velocities are measured along an ultrasonic beam (shown in figure~\ref{fig:hat}) analysing the Doppler shift effect of reflected ultrasonic waves \citep{Bri01}. Figure~\ref{fig:DTO} show the spatio-temporal diagrams of the velocity measured along the beam. The middle of the beam is located near the inner core (at about $1.7$cm) whereas both ends are near the outer sphere. The top of the vertical axis in figure~\ref{fig:DTO} corresponds to the ultrasonic source. The measured velocity is the projection of the velocity field along the beam and thus corresponds to the azimuthal velocity near the inner core. Note that for a centripetal radial velocity, the measured velocity is positive close to the ultrasonic source and negative close to the beam end (bottom of the vertical axis in figure~\ref{fig:DTO}). A periodic pattern and a temporal drift are visible in the spatio-temporal diagrams. Since the velocity measurement is located $2.5$ cm below the equatorial plane, the periodic increase of the velocity and its temporal drift shows the travelling of the $m=2$ azimuthal wave instabilities. Smaller structures corresponds to higher azimuthal numbers. For flow approximatively $30$ times critical (figure~\ref{fig:DTO}(b)), the linear mode is still clearly visible even if higher modes strongly imprint the flow. The spatial structure of the oscillation and the temporal period are found to be the same in the experiment and in the numerical simulation for the Reynolds numbers investigated (from $7200$ to $31678$) corresponding to the lowest range of Reynolds numbers experimentally reachable. By contrast, due to the turbulent behaviour of the flow at these Reynolds numbers, the numerical simulations require high resolution. For instance, the calculation for $Re=31678$ requires $300$ radial grid points and truncated spherical harmonic degree $l=300$ and order $m=48$. The agreement between numerical and experimental results is impressive and validates the numerical code.

\subsection{Dynamo instability}
\label{sec:magE0}
\begin{figure}
\begin{center}
	\resizebox*{12cm}{!}{\includegraphics{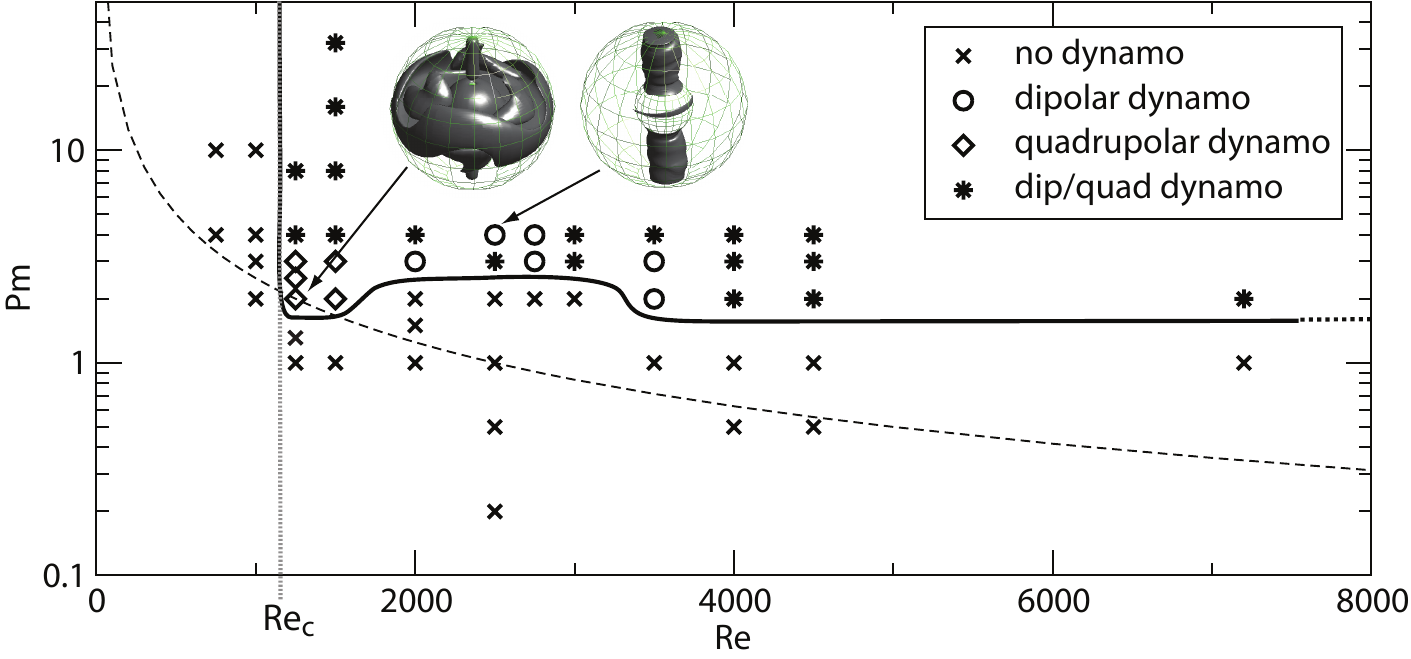}}
	\caption{Dynamo simulations in the parameter space $Pm$ versus $Re$ without global rotation ($1/E=0$). The circles, diamonds and stars represent the simulations that lead to dynamos. The circles are dipolar dynamos, the diamonds are quadrupolar dynamos and the stars are mixed dipolar/quadrupolar dynamos. The crosses are failed dynamos. The solid thick line symbolizes the dynamo onset. The dashed black line represents the value $Rm=2500$. The vertical dotted gray line represents the value $Re_c$ at which the first non-axisymmetric hydrodynamical instability occurs. An isolevel of magnetic energy corresponding to $30\%$ of its maximum value is drawn at the top of the plot. The arrows indicate the parameters used in the simulation: $Re=1250$, $Pm=2$ (snapshot at the peak of magnetic energy) (left) and $Re=2500$, $Pm=4$ (snapshot) (right).}
	\label{fig:PmReE0}
\end{center}
\end{figure}

For a given Reynolds number $Re$ we find the dynamo threshold by varying the magnetic Prandtl number $Pm$ in equation (\ref{eq:ind}). The dynamo onset is determined by following the time evolution of the magnetic energy over several global magnetic diffusion times. The dynamo region of parameter space is presented in figure~\ref{fig:PmReE0}. A large range of Reynolds numbers have been investigated (up to $Re=7200$). We have found dynamos for magnetic Reynolds numbers $Rm > 2500$ (and magnetic Prandtl numbers greater than 1). The dynamos appear for $Re>Re_c$: the dynamo onset occurs when the hydrodynamical undular instability of the equatorial jet described in the previous section is present. Therefore we conclude that the dynamo effect arises from the non-axisymmetric flow associated with this instability.
\newline \indent
Our results can be compared with those of \citet{Dud89}, who performed numerical simulations of kinematic dynamos in spherical geometry. In particular, they studied the growth rate of a magnetic field generated by a $\mathbf{t}_{10} +\epsilon \mathbf{s}_{20}$ flow where $\mathbf{t}$ and $\mathbf{s}$ are the toroidal and poloidal vectors respectively, the first subscript denotes the degree and the second subscript the mode of the spherical harmonics. The flow consists of two axisymmetric rolls, one in each hemisphere, mirror symmetric about the equatorial plane. Our axisymmetric flow geometry is thus similar to this flow. With an appropriate $\epsilon$, the poloidal to toroidal flow ratio, Dudley \& James found that this flow can sustain a magnetic field provided that the meridional flow $\mathbf{s}_{20}$ is directed inward in the equatorial plane and outward near the polar axis. Surprisingly with a meridional flow of opposite sign (i.e. a centrifugal flow as in our configuration) they did not find a dynamo threshold. As previously discussed, we did not find dynamos with a purely axisymmetric flow, in agreement with the work of Dudley \& James.
\newline \indent 
In cylindrical Couette flow, dynamo magnetic fields can be excited by axisymmetric Taylor vortices \citep{Lau00,Wil02}. In spherical geometry, the presence of Taylor vortices is restricted to the small gap problem \citep{Khl68}. Nonetheless, a pair of Taylor vortices in the cylindrical case is fairly similar to a pair of spherical axisymmetric Couette rolls (for a graphical representation see the meridional circulation streamlines plotted in the upper hemisphere in figure~\ref{fig:hat}). Whereas the former flow can sustain a magnetic field, we found that the latter can not. Possibly the $z$-periodicity of the Taylor-Couette rolls in the cylindrical case plays a major role in the dynamo mechanism. In particular it might be related to the boundary conditions for the velocity $\vel$ and the electric currents $\cur$. For instance a shift of phase between $\vel$ and $\cur$ is hard to achieve in the spherical case as they have to match the boundary conditions at the ``top'' and ``bottom'' of the spherical container ($\vel=\mathbf{0}$ and $\cur=\mathbf{0}$) contrary to the cylindrical problem where no such boundary conditions are imposed at the top and bottom of the periodic box.
\newline \indent
We observe from figure~\ref{fig:PmReE0} that increasing the Reynolds number, and thus the flow fluctuations, does not affect the $Pm$ dynamo threshold, which remains around a constant critical magnetic Prandtl number $Pm_c \sim 2$. Accordingly we infer that, whereas the first instability of the flow triggers the instability of the self-induced magnetic field, secondary fluctuations produced by increasing the boundary forcing do not assist the dynamo mechanism.
\newline \indent
The structure of the self-sustained magnetic field varies considerably between different regions of parameter space. We first report our results in the regime close to the $Re$ dynamo onset ($Re\le 1500$) and then in the
supercritical regime ($Re>1500$).

\subsubsection{Magnetic field for $Re\le 1500$}
\label{sec:magE0_a}
\begin{figure}
\begin{center}
	\resizebox*{12cm}{!}{\includegraphics{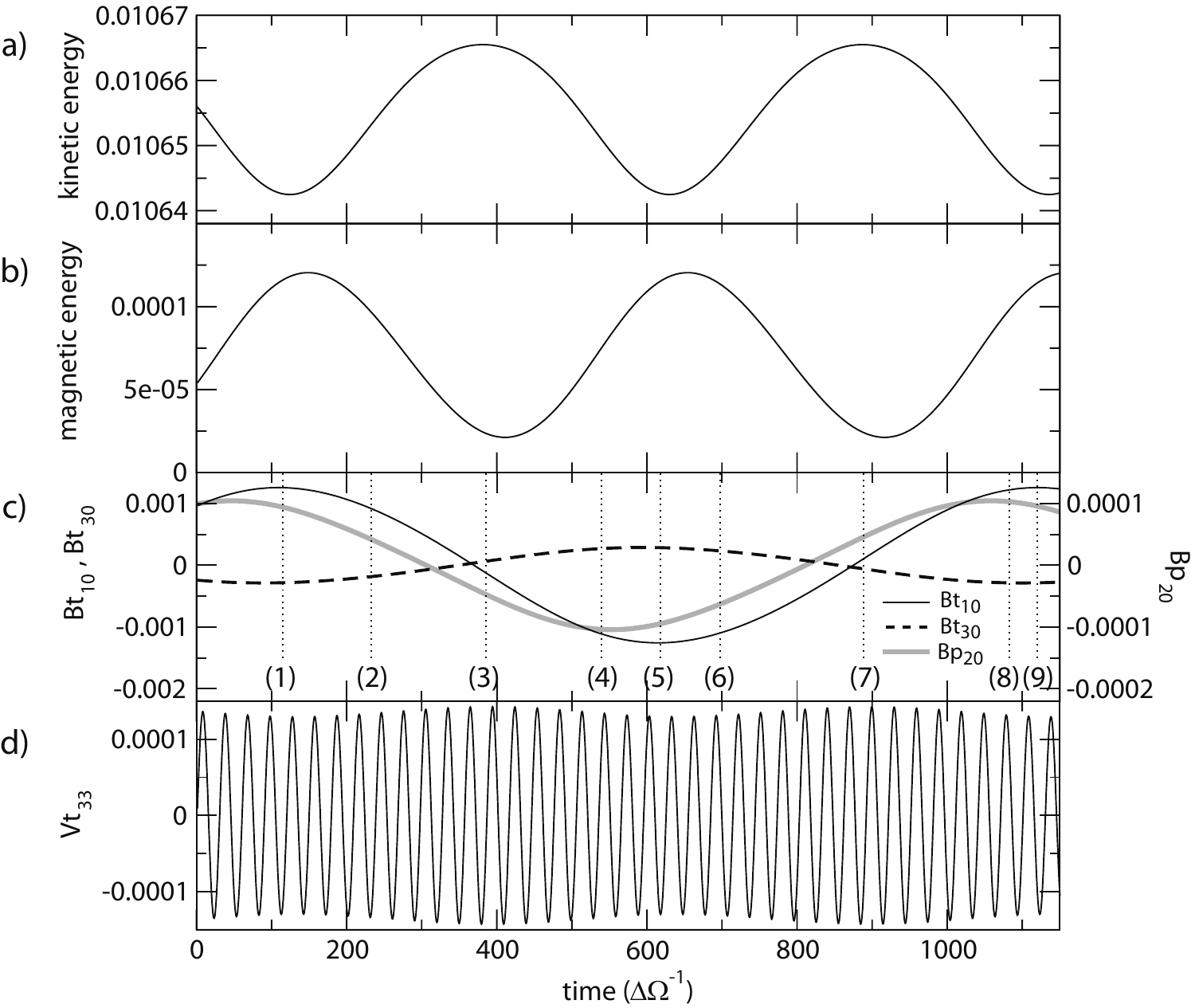}}
	\caption{Time series for (a) the kinetic energy, (b) the magnetic energy, (c) the magnetic toroidal scalars $Bt_{10}$ (black line) and $Bt_{30}$ (dashed black line) (left vertical axis) and poloidal scalar $Bp_{20}$ (thick gray line) (right vertical axis) and (d) the flow toroidal scalar $Vt_{33}$. The parameter of the simulation are $1/E=0$, $Re=1250$ and $Pm=2$. The vertical dotted black lines on (c) indicated the time for each snapshots (from (1) to (9)) of figure~\ref{fig:dyn_osc2}. The energies are averaged over the whole shell. The poloidal and toroidal scalars are the values at a radius $r=0.995$.}
	\label{fig:dyn_osc}
\end{center}
\end{figure}

\begin{figure}
\begin{center}
	\resizebox*{12cm}{!}{\includegraphics{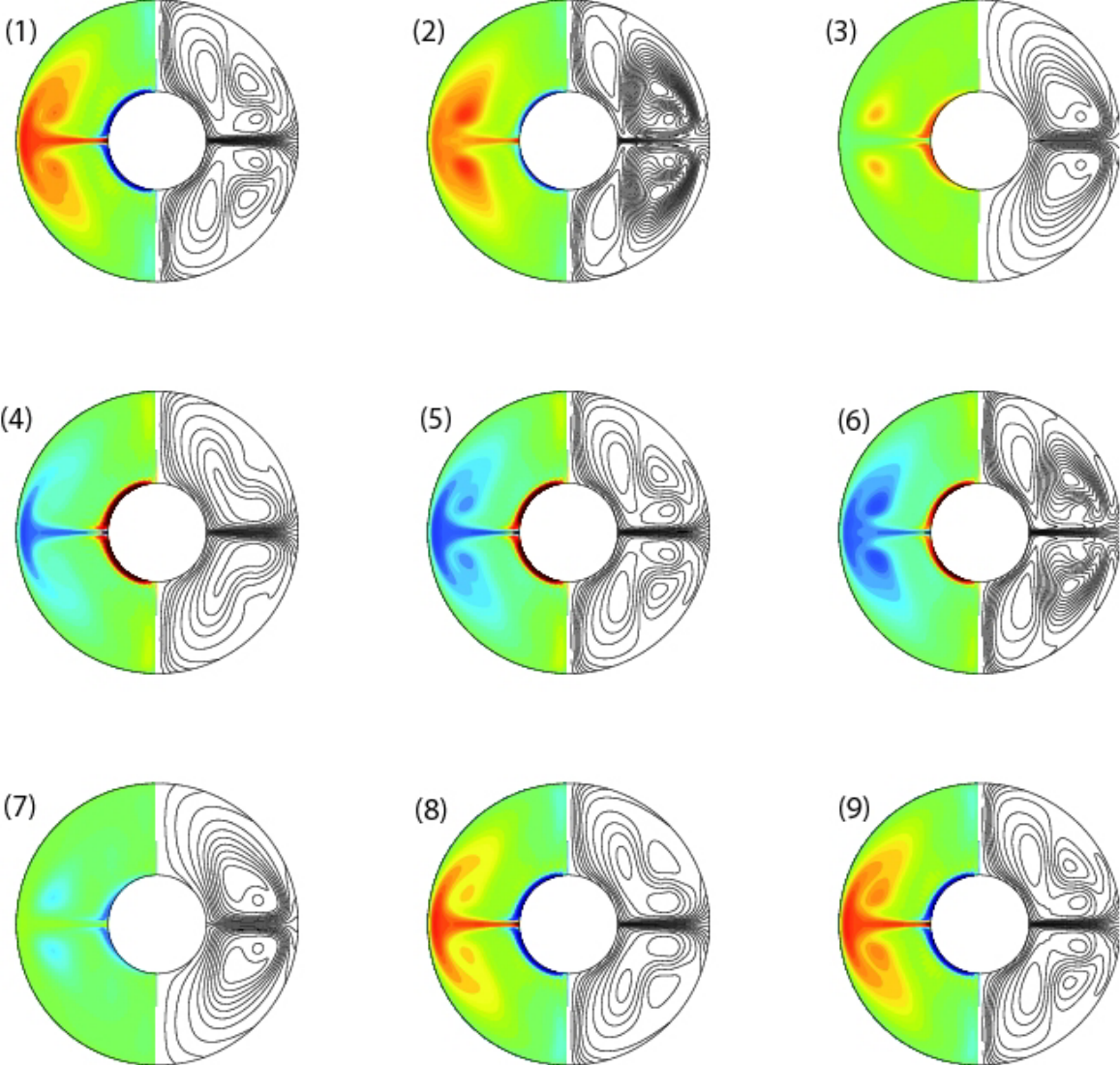}}
	\caption{Snapshots of the intensity of the axisymmetric azimuthal magnetic field (left) and axisymmetric poloidal magnetic field lines (right) in a meridional plane. The color scale of the toroidal field intensity is the same for all the plots. The time of each snapshot ((1) to (9)) is indicated in figure~\ref{fig:dyn_osc}(c) by the vertical dotted black lines.}
	\label{fig:dyn_osc2}
\end{center}
\end{figure}

\ \\
Close to the dynamo threshold we expect the magnetic field to exhibit a dipole-quadrupole decoupling because of selection rules \citep{Dud89}.
Near the critical Reynolds number ($Re<1.6Re_c$) and the critical magnetic Prandtl number ($Pm\le 1.5Pm_c$), we found quadrupolar dynamos with a dominant axial quadrupole that oscillates regularly (diamonds on figure~\ref{fig:PmReE0}). The magnetic and kinetic energies oscillate out of phase (figure~\ref{fig:dyn_osc}(a) and (b)). We note that the phase shift is not exactly half a period: the kinetic extrema are slightly in advance with respect to the magnetic ones. The period of the oscillation is about $0.07\tau_{\eta}$ where $\tau_{\eta}=\frac{r_o^2}{\eta}$ is the global magnetic diffusion time.
The axial quadrupole (denoted $Bp_{20}$) and the axisymmetric toroidal scalars of degree $l=1$ and $l=3$ (denoted $Bt_{10}$ and $Bt_{30}$, respectively) are the dominant magnetic components and oscillate around a zero-mean value (figure~\ref{fig:dyn_osc}(c)). The axial dipole is much weaker and decays to zero over the course of the simulation.
If we suppress the Lorentz force in the simulation after a given time, so that the time evolution of the velocity field is decoupled from that of the magnetic field, then the kinetic energy returns to the (stationary)
value found in the purely hydrodynamical case. However, the magnetic energy continues to oscillate regularly while increasing exponentially (not shown). As the oscillation survives only in the magnetic field when it stops acting on the flow, the oscillation is ascribed to the magnetic field. The kinetic energy oscillation occurs only when the Lorentz force is present and thus is slaved to the magnetic oscillation.
For a frozen flow i.e. a steady flow taken from an arbitrary moment during the simulation, the magnetic field decays to zero. The time-dependence of the velocity field is necessary to sustain the oscillating quadrupolar magnetic field. As described in section~\ref{sec:hydroE0}, a non-axisymmetric wave propagates in the azimuthal direction when $Re$ is greater than a critical value (around $1100$). For the Reynolds numbers investigated in this section ($Re=1250$ and $Re=1500$), the most unstable mode is $m=3$. The signature of the wave is visible in the time series of $Vt_{33}$, the $m=3$ toroidal scalar of degree $l=3$ (figure~\ref{fig:dyn_osc}(d)). This wave is of higher frequency than the magnetic oscillation. This hydrodynamical wave is both non-axisymmetric and time dependent. We have found that both these ingredients are necessary to sustain oscillating quadrupolar dynamos, and we conjecture that the non-axisymmetric wave is central to the dynamo mechanism. Figure~\ref{fig:dyn_osc2} shows snapshots of the intensity of the axisymmetric toroidal magnetic field and the axisymmetric poloidal magnetic field lines. During a cycle the toroidal magnetic field is generated in the equatorial jet and around the inner core where the velocity shear is largest. The toroidal magnetic field is likely created by the shearing of poloidal magnetic field lines, i.e. by the $\omega$-effect. Cells of poloidal magnetic field form close to the outer sphere and propagate toward the axis. It is tempting to invoke the so-called $\alpha$-effect to explain the generation of the poloidal field via the interaction between the $m=3$ hydrodynamical wave and the magnetic field.

\begin{figure}
\begin{center}
	\resizebox*{12cm}{!}{\includegraphics{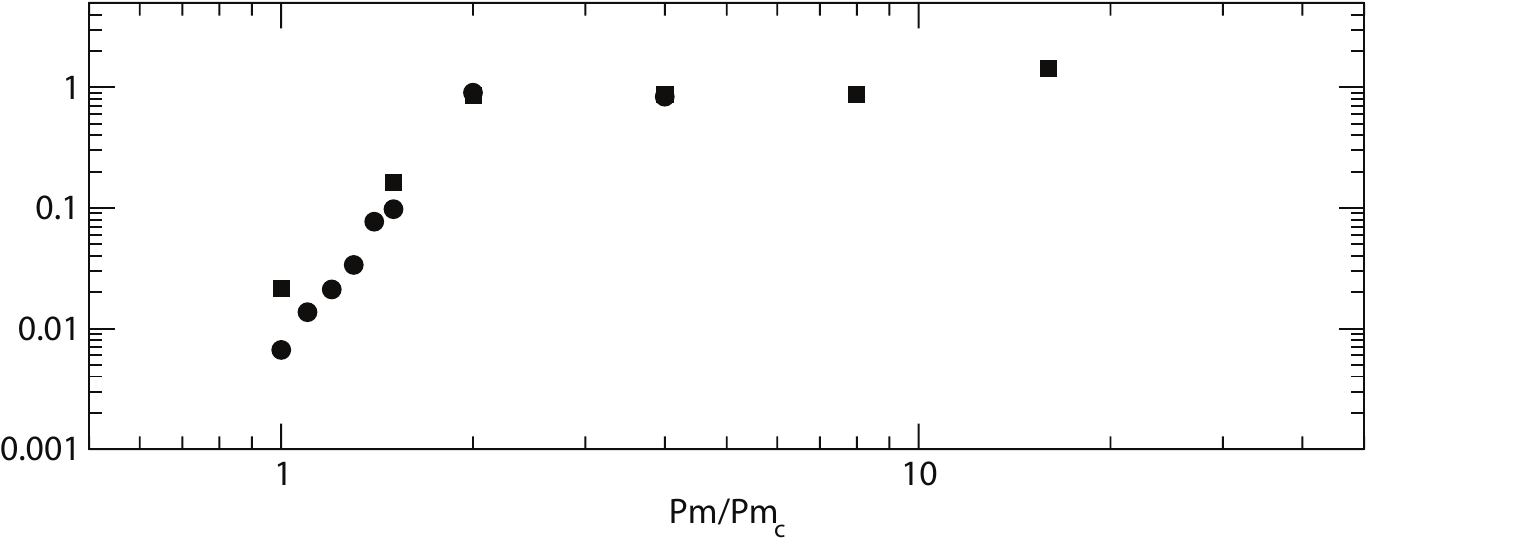}}
	\caption{Ratio of the temporally averaged values of the magnetic energy of the fluid to the kinetic energy in function of $Pm/Pm_c$ for the parameters $1/E=0$, $Re=1250$ (circles) and $Re=1500$ (squares). The black circles and squares for $Pm<2Pm_c$ represents the oscillating quadrupolar dynamos. For $Pm>2Pm_c$ the magnetic field is a time fluctuating mixed dipolar/quadrupolar field.}
	\label{fig:E0satur}
\end{center}
\end{figure}

\indent
A slight increase of the forcing (from $Re=1250$ to $Re=1500$) produces a lengthening of the (dimensionless) oscillation period and amplitude for both kinetic and magnetic components. An increase of $Pm$ (up to $Pm=3$) produces the same effect. However, for $Pm>3$ or $Re>1500$ the time oscillating behaviour disappears and we obtain a more complex magnetic field with both dipole and quadrupole components with time fluctuating intensities (stars on figure~\ref{fig:PmReE0}). The dynamo field in this case is stronger than the oscillating quadrupole described above. Figure~\ref{fig:E0satur} shows the ratio of magnetic energy to kinetic energy for moderate values of $Re$ ($Re< 1.6Re_c$) as $Pm$ increases. Equipartition between magnetic and kinetic energy density is achieved when increasing $Pm$ above twice its critical value where the dynamo is a mixed dipole/quadrupole field. The magnetic oscillation is confined to the weak magnetic field regime ($Pm<2Pm_c$).

\subsubsection{Magnetic field for $Re>1500$}
\begin{figure}
\begin{center}
	\resizebox*{15cm}{!}{\includegraphics{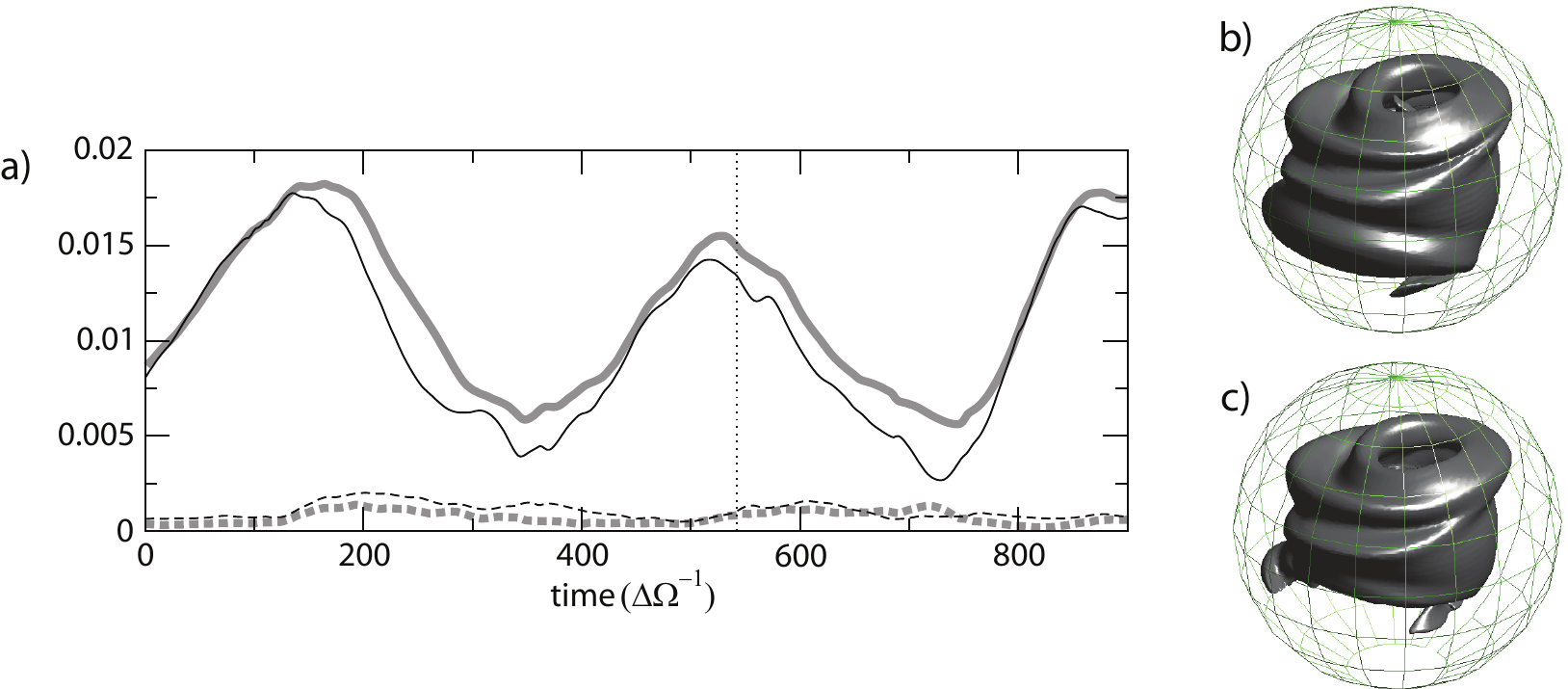}}
	\caption{(a) Time evolution of the magnetic energy (thin black line) and  the kinetic energy (thick gray line) for the simulation $1/E=0$, $Re=2500$ and $Pm=3$. The solid line represents the toroidal part and the dashed line the poloidal part. The energies are dimensionless; the scaling factor is $\rho (r_i \Delta \Omega)^2$. (b) An isolevel of the magnetic energy and (c) of the kinetic energy corresponding to $20\%$ of the maximum value at the time represented by the dotted line in figure (a).}
	\label{fig:dyn_fluc}
\end{center}
\end{figure}

\ \\
Further from the dynamo threshold (above $Pm=3$ or $Re=1500$) the magnetic field is either mainly an axial dipole (circles on figure~\ref{fig:PmReE0}) or a complex field with time-dependent dipolar and quadrupolar components (stars on figure~\ref{fig:PmReE0}). The former field is either stationary or fluctuating with reversals with no regular period. In the latter case, the system undergoes large fluctuations (figure~\ref{fig:dyn_fluc}(a)). Chaotic reversals occur and the predominance of the dipole or the quadrupole families varies over time. An important feature of these dynamos is that equipartition between magnetic and kinetic energies is achieved even close to the $Pm_c$ dynamo threshold, as one can notice in figure~\ref{fig:dyn_fluc}(a).

As expected for Couette flows which create a strong azimuthal velocity component compared to the meridional circulation, the toroidal magnetic field is stronger than the poloidal field (about $2$ to $10$ times stronger). The magnetic field is strongest around the inner core where the azimuthal shear flow is largest (figure~\ref{fig:dyn_fluc}(b)). The interaction parameter which measures the relative importance of Lorentz forces to inertia is of the order of unity as the equipartition of energy is achieved. As a result the saturated magnetic field has a strong feedback on the flow due to Lorentz forces and we expect the magnetic fluctuations to be recorded by the flow. Indeed the flow is no longer jet-like, and is located where the magnetic field is present (figure~\ref{fig:dyn_fluc}(c)). Imposing a frozen flow does not kill the dynamo contrary to the oscillating quadrupolar dynamos found close to the dynamo onset ($Re\le 1500$, $Pm\le 3$). Consequently the time fluctuations of the flow are not vital for the dynamo mechanism in the case of supercriticality ($Re>1500$). This is consistent with the fact that increasing $Re$, and thus the flow fluctuations does not modify the $Pm$ threshold (figure~\ref{fig:PmReE0}). To summarize, close to the hydrodynamical threshold ($1250\le Re \le 1500$), the (undular) non-axisymmetric hydrodynamical oscillation is crucial to trigger the dynamo instability. However in the supercritical case ($Re>1500$) the small scale fluctuations due the increase of the mechanical forcing are powerless for the dynamo process but are not fatal either.
\newline \indent During a simulation, when decreasing the Reynolds number from $Re>Re_c$ to $Re<Re_c$ for a fixed magnetic Prandtl number $Pm$, one can expect to maintain the dynamo and obtain a so-called subcritical dynamo. The case with $Pm$ large enough in order to reach the equipartition of energy may be a good candidate. In this case we did not find any subcritical dynamo. When decreasing the Prandtl number from $Pm>Pm_c$ to $Pm<Pm_c$ at a fixed Reynolds number $Re>Re_c$, no subcritical dynamo has been found either.

\section{Influence of the outer sphere rotation}
\label{sec:E}
The Coriolis force acting on the system is large when the Ekman number $E$ is small. According to the Proudman-Taylor theorem, we expect a strong change in the flow compared to the previous case if the Ekman number is small enough. Decreasing $E$ requires high resolution since the viscous boundary layers scale as $E^{1/2}$. We compute simulations with $E$ of the order of $10^{-3}-10^{-4}$ which is already very demanding whereas in most natural systems the Ekman number is much smaller (for instance $E\le 10^{-14}$ in the Earth's core (based on metal liquid viscosity)).

\subsection{Hydrodynamics}
\begin{figure}
\begin{center}
	\resizebox*{12cm}{!}{\includegraphics{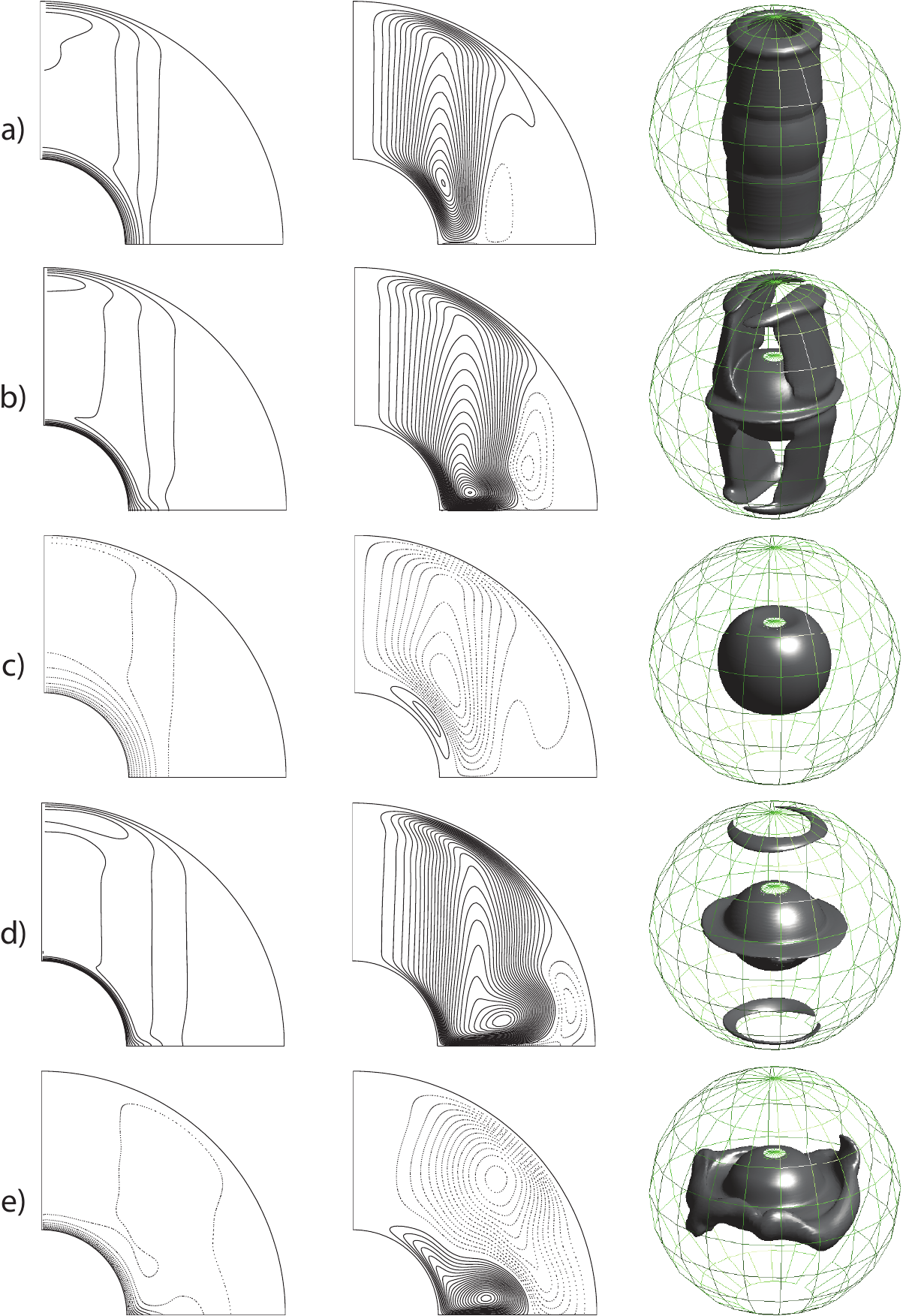}}
	\caption{Contours of the angular velocity (zonally averaged) with a contour interval of $0.1$ (left), streamlines of the meridional circulation with a contour interval of $1.6\times10^{-4}$ (middle) and isolevel of the kinetic energy corresponding to $25\%$ of the maximum value (right) for (a) $Ro=0.01$, (b) $Ro=1$, (c) $Ro=-1$, (d) $Ro=2.5$ and (e) $Ro=-2.5$. All these simulations are purely hydrodynamic and for $E=10^{-3}$. The black lines represents the positive values and the dashed lines the negative values of the angular velocity (left) and of the streamfunction associated to the meridional flow (middle).}
	\label{fig:flow_E3}
\end{center}
\end{figure}

\begin{figure}
\begin{center}
	\resizebox*{12cm}{!}{\includegraphics{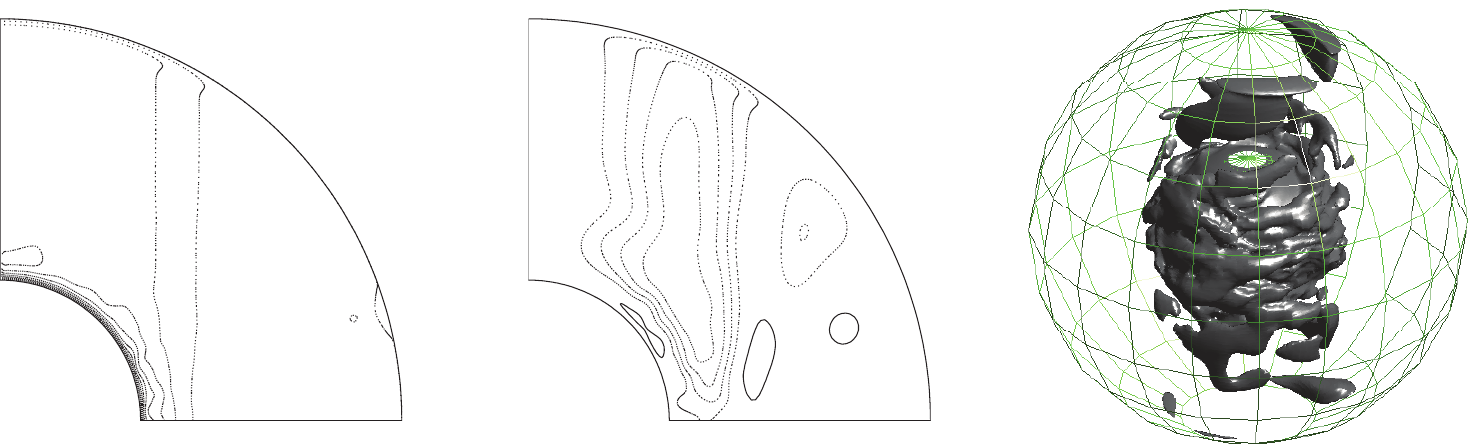}}
	\caption{Same representation than figure~\ref{fig:flow_E3} for $Ro=-1$, $E=10^{-4}$ and $Pm=0$.}
	\label{fig:flow_E4}
\end{center}
\end{figure}

In the basic state with a rapid overall rotation, a small differential rotation generates a Stewartson layer in the bulk of the fluid. This layer is well understood and has been studied theoretically \citep{Ste66}, numerically \citep{Dor98,Hol03} and experimentally \citep{Hid67,Sch05}. It consists mainly on an axisymmetric layer of size $E^{1/4}$ accommodating the jump of angular velocity between the fluid inside and outside the tangent cylinder i.e. the cylinder parallel to the axis of rotation and touching the inner core at the equator. The Stewartson layer becomes unstable for a critical shear imposed by a differential rotation of the inner core measured by the Rossby number $Ro=ReE$, as studied numerically by \citet{Hol03} with fully three-dimensional simulations ($E=10^{-4}$ to $E=10^{-5}$) and by \citet{Sch05} with a quasi-geostrophic (QG) model ($E=10^{-5}$ to $E=10^{-10}$). In a spherical shell they found that non-axisymmetric instabilities develop for a critical Rossby number $Ro_c \propto E^{1/2}$. According to Schaeffer \& Cardin, the critical instability responsible for the destabilization of the Stewartson layer is a Rossby wave. Experiments have been carried out in spherical shell geometries \citep{Hol04,Sch05}. The agreement between experimental and numerical results concerning the critical Rossby number, the most unstable mode and the wave frequency is striking in both cases. The non-linear regime has been studied numerically by \citet{Sch05b} using their quasi-geostrophic approach leading to a peculiar Rossby wave turbulence (for $E=10^{-6}$ to $E=10^{-8}$ and $|Ro|=0.1$ corresponding to $Ro\sim 30 Ro_c$). Recently experiments in a $60$-cm diameter spherical shell with differential rotation have been carried out by \citet{Kel07} using liquid sodium as the working fluid and magnetic induction as a probe for the flow field in the non-linear regime (down to $E \sim 10^{-8}$ and up to $|Ro| \sim 0.1$). Instead of Rossby waves instabilities, they show the presence of inertial modes. To explain the apparent discrepancy between the results of \citet{Sch05b} and \citet{Kel07}, we may conjecture that as Kelley \emph{et al.} measured the radial component of the magnetic field $B_r$ and as their imposed magnetic field is an axial dipole, they can not probe the presence of Rossby waves in their experiment. Indeed according to the frozen-flux radial induction equation at the outer boundary, a geostrophic motion does not induce a radial magnetic field if $B_r\propto \cos\theta$ \citep[see][]{LeM84,Pai08}.
\newline \indent
Our purely hydrodynamical simulations are far from being in the asymptotic limit $E\to 0$ ($E\ge10^{-4}$ in our simulations). A direct comparison between our results with the theory is worthless as \citet{Dor98} shows that the asymptotic regime can not be illustrated numerically for $E>10^{-5}$ as the Stewartson layer scales as $E^{1/4}$. Our goal is to understand the instabilities arising from an increase in $Ro=ReE$ largely above the hydrodynamical instability onset in order to relate them to the dynamo instability when the fluid is conducting. The Rossby number $Ro$ is a measure of the ratio of the inertial forces to the Coriolis force and is then the relevant dimensionless number in our study, since the balance between inertial forces and Coriolis force is the key ingredient for the flow dynamics. A positive value for $Ro$ means that the inner sphere rotates faster than the outer sphere whereas a negative $Ro$ means that the inner core rotates slower than the outer sphere or in counter-rotation. We determine different regimes depending on the value of $Ro$:
\begin{itemize}
 \item When $|Ro|$ is less than $1$ (figure~\ref{fig:flow_E3}(a)), the Coriolis force is predominant and the flow is roughly invariant in the rotation direction. As expected a cylindrical Stewartson layer develops. For a positive $Ro$ the meridional circulation linked to the Ekman pumping on the outer boundary is counter-clockwise in the upper right half of the domain. It is a clockwise meridional flow for a negative $Ro$.
 \item When $|Ro|$ is about 1, the inertial terms balance the Coriolis force and the Proudman-Taylor constraint vanishes. The equatorial jet becomes stronger and destabilizes the Stewartson layer. For a positive $Ro$ (figure~\ref{fig:flow_E3}(b)), the flow remains nevertheless roughly $z$-invariant. The instability is a Rossby wave with an azimuthal wavenumber $m=2$ propagating in the prograde direction as expected with the sign of the slope. For a negative $Ro$ (figure~\ref{fig:flow_E3}(c)) the meridional circulation linked to the Ekman pumping and the centrifugal flow are in opposite direction leading to a strong shear layer at the inner core boundary. This flow is not $z$-invariant. $|Ro| \sim 1$ corresponds to a transition between quasi-geostrophic (nearly 2D) turbulence to 3D turbulence. As a consequence, the decrease in the Ekman number $E$ (see figure~\ref{fig:flow_E4} for $E=10^{-4}$) leads to the computation of simulations at high Reynolds number displaying small scale 3D fluctuations. The shear layer present when $Ro\sim-1$ and $E=10^{-3}$ becomes unstable and partially disappears to the benefit of small scale structures when $E=10^{-4}$.
 \item When $|Ro|$ is greater than $1$ the inertial forces are predominant and the equatorial jet is strengthened. 3D small scale fluctuations appear. For a positive $Ro$ (figure~\ref{fig:flow_E3}(d)), the Rossby waves vanishes and the flow is mainly concentrated in the equatorial jet. For a negative $Ro$ (figure~\ref{fig:flow_E3}(e)) the shear layer loses its intensity and a jet-like flow is now present. The flow for large absolute Rossby numbers becomes more and more similar to the non-rotating case.
\end{itemize}

\subsection{Dynamo instability}
\begin{figure}[h]
\begin{center}
	\resizebox*{12cm}{!}{\includegraphics{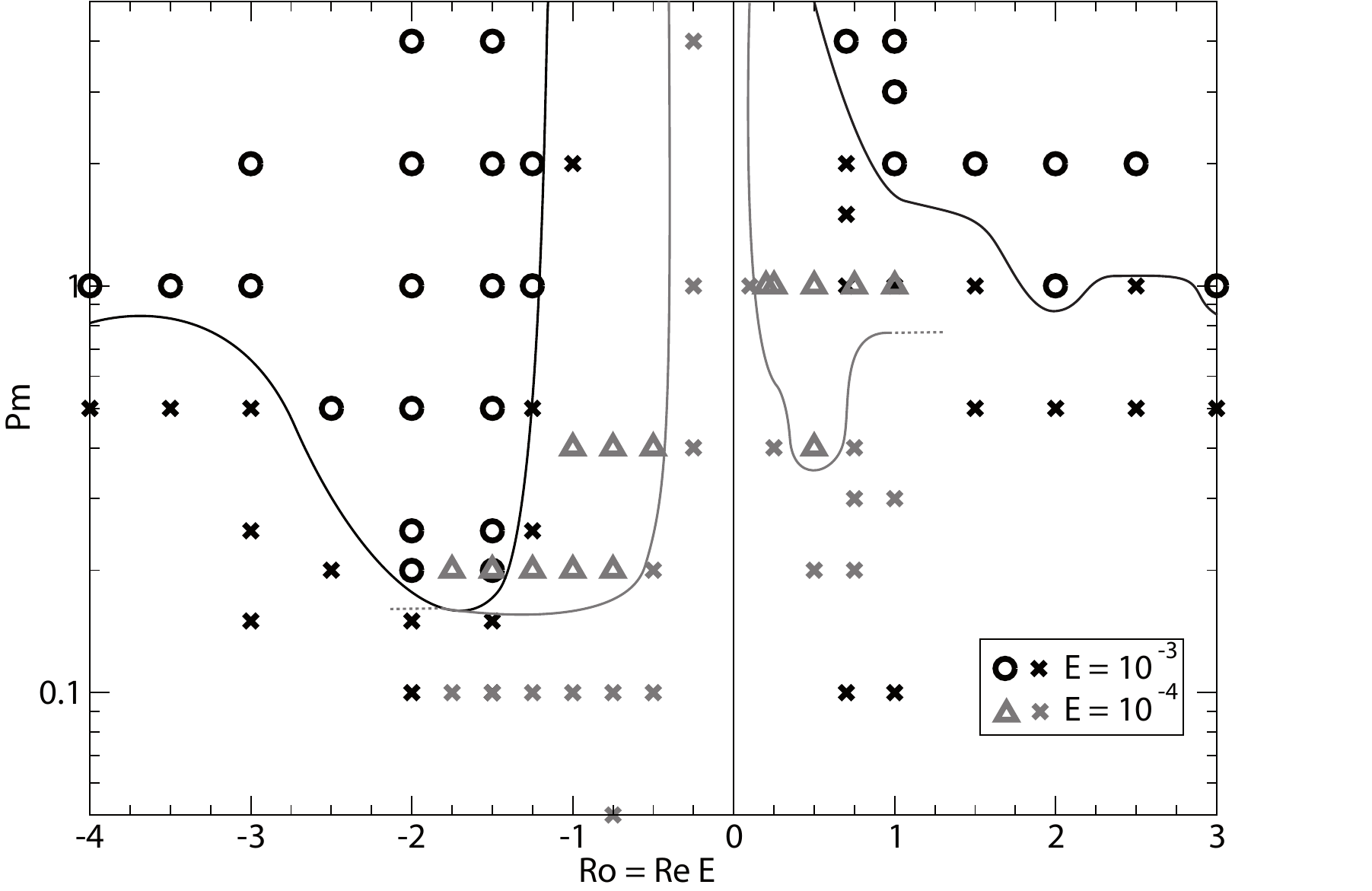}}
	\caption{Dynamo diagram in the parameter space $Pm$ versus $Ro$ for $E=10^{-3}$ (black circles) and $E=10^{-4}$ (gray triangles). The cross are failed dynamos. The black and gray lines represent the dynamo onset for $E=10^{-3}$ and $E=10^{-4}$ respectively. The hydrodynamical onset at which the nearly 2D flow (Stewartson layer) is unstable is for $E=10^{-3}$: $Ro^-_c \in [-1.5,-1]$, $Ro^+_c \in [0.316,0.666]$ and for $E=10^{-4}$: $Ro^-_c \in [-0.25,-0.1]$, $Ro^+_c \in [0.1,0.2]$.}
	\label{fig:PmRoE}
\end{center}
\end{figure}

As already mentioned, few studies have been carried out on the dynamo generation by a shear layer in a spherical geometry with an overall rotation. \citet{Sch06} computed dynamos generated in a split-sphere geometry by a quasi-geostrophic flow. They found dynamos at low magnetic Prandtl number $Pm$ when decreasing the Ekman number $E$ (down to $Pm=0.003$ for $E=10^{-8}$). The QG assumption allows them to compute rapid rotation (from $E=10^{-6}$ to $E=10^{-8}$). They also found that increasing the mechanical forcing ($Ro$) for a given $E$ reduces the critical magnetic Prandtl number $Pm_c$. The critical magnetic Reynolds number $Rm_c$ is roughly constant (of the order of a few thousand) and independent of $E$ and $Pm$.
\newline \indent
Figure~\ref{fig:PmRoE} presents the dynamo diagram of our simulations worked out at $E=10^{-3}$ and $E=10^{-4}$. A strong disymmetry occurs with the sign of the Rossby number $Ro$: the negative differential rotation case lowers the dynamo onset to $Pm_c=0.2$ and $Rm_c=300$ when $E=10^{-3}$. When $E=10^{-4}$, the critical magnetic Prandtl number is also $Pm_c=0.2$ but $Rm_c$ equals $1500$ for negative differential rotation. Unlike the results of \citet{Chr06} with convective dynamos and of \citet{Sch06} discussed previously, $Rm_c$ does not remain constant when decreasing $E$. An explanation of this result is given by the study of the flow dynamics in the previous section: the shear layer present around the inner core when $Ro\sim-1$ and $E=10^{-3}$ (figure~\ref{fig:flow_E3}(c)) is likely to be very efficient for dynamo action as it shears the poloidal magnetic field lines and creates a strong toroidal magnetic field. Thanks to this mechanism, a ``dynamo window'' is present at relatively low $Rm$ for $E=10^{-3}$ and $-1.5\ge Ro\ge -2$. The destabilization of this shear layer by the small scale 3D fluctuations when decreasing $E$ (figure~\ref{fig:flow_E4}) leads to the disappearance of the dynamo window and therefore the requirement of higher $Rm$ for dynamo action. 

\begin{figure}
\begin{center}
	\resizebox*{15cm}{!}{\includegraphics{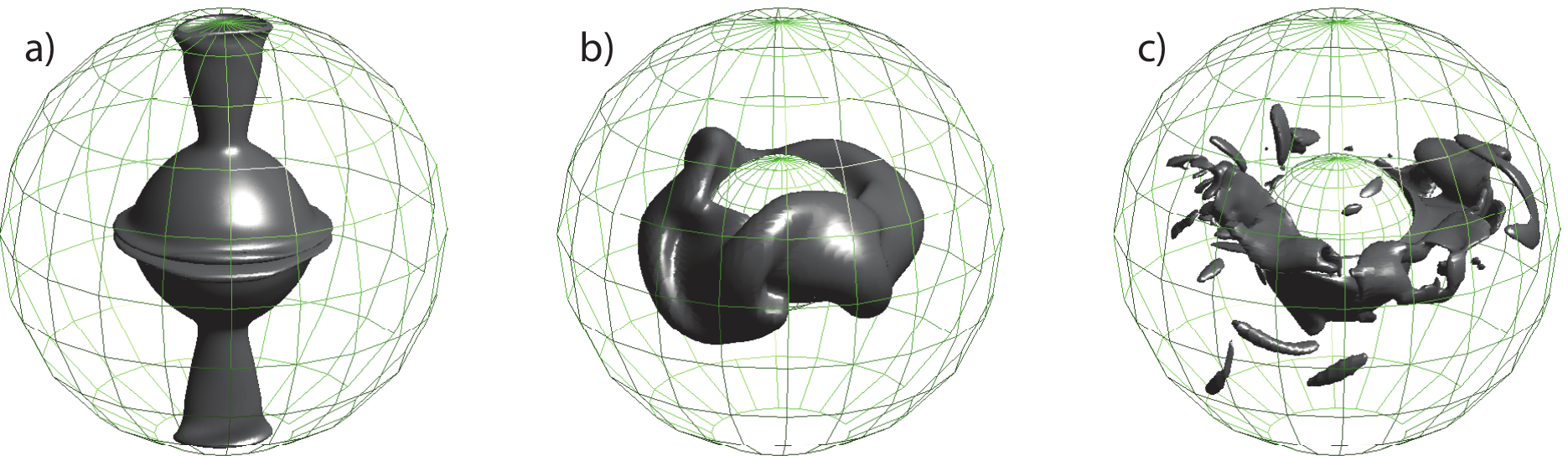}}
	\caption{An isolevel of the magnetic energy correponding to $25\%$ of the maximal value for $E=10^{-3}$ for dynamo fields close to the threshold: (a) $Re=1000$, $Pm=2$, (b) $Re=-1500$, $Pm=0.2$ (middle) and (c) $Re=-3500$, $Pm=1$.}
	\label{fig:ME_E3}
\end{center}
\end{figure}

In the DTS experiment, \citet{Nat08} reported an interesting observation with regards to the results previously discussed. The DTS setup in Grenoble is a spherical Couette experiment of $42$-cm diameter using liquid sodium as working fluid and with an imposed dipolar magnetic fluid. \citet{Nat08} showed that a peak of the induced magnetic field occurs in the counter-rotation case corresponding to the regime where the fluid rotation rate is almost equal and opposite to the outer sphere rotation rate ($E\sim 10^{-7}$). According to \citet{Nat08} the appearance of this strong induced field is due to the fact that the fluid is almost at rest in the laboratory frame and so the meridional flow is free from the Proudman-Taylor constraint and becomes stronger. As they suggested, dynamo action may be favored in this situation. It is thus possible than other ``dynamo windows'' related to peculiar hydrodynamical regimes exist at low Ekman numbers. 
\newline \indent
\citet{Sch06} found that $Pm_c$ depends on the sign of $Ro$ with negative differential rotation yielding a slightly lower dynamo threshold. In their work the Ekman numbers are at least $2$ orders of magnitude smaller than in this study and their QG assumption restricts the exploration of the non-linear regime to Rossby numbers smaller than $1$ \citep[up to $0.1$ in][]{Sch06}. Our strong dependence on the sign of $Ro$ when $E=10^{-3}$ is due to the presence of the strong ageostrophic shear layer around the inner core for negative $Ro$. The weakening of this shear layer when $E=10^{-4}$ yields a less marked dependence on the sign of $Ro$.
\newline \indent
When $|Ro|$ increases ($Ro<-2$) for $E=10^{-3}$, the shear layer disappears and therefore the generation of toroidal magnetic field is less important. This could explains why we find dynamos only at higher $Pm$. 
\newline \indent 
We emphazise that similarly to the case without global rotation, the first non-axisymmetric hydrodynamic instability (occuring at $Ro_c$) is necessary to generate a dynamo when a global rotation is imposed. Moreover the dynamo onset occurs for lower $Ro$ when decreasing $E$. These results are consistent with the results of \citet{Hol03} and \citet{Sch05} in the hydrodynamical case who find that the critical Rossby number $Ro_c$ decreases with $E$ (and in our case triggers the dynamo instability). Here again if we except the dynamo window for $E=10^{-3}$ ($-1.5 \ge Ro \ge -2$) we notice that the $Pm_c$ dynamo threshold remains roughly flat when increasing the Reynolds number. Consequently the flow fluctuations are powerless to lower the threshold for either positive or negative differential rotation but are not fatal to the dynamo action. The same result has been shown without global rotation (section \ref{sec:magE0}).
\newline \indent 
An isolevel of the magnetic energy is shown in figure~\ref{fig:ME_E3} for different $Re$ at the dynamo onset ($Pm_c$). When $Re>0$ (figure~\ref{fig:ME_E3}(a)) the magnetic field is mainly an axial dipole. The magnetic energy is located around the inner core and close to the rotation axis inside the tangent cylinder. In the dynamo window at $Ro\sim-1.5$ (figure~\ref{fig:ME_E3}(b)) a strong toroidal magnetic field is created in the shear layer and twisted around the inner core. When the flow becomes more turbulent (figure~\ref{fig:ME_E3}(c)) the magnetic field is scattered close to the equatorial jet. The time evolution of the magnetic and kinetic energy become rapidly chaotic above the dynamo threshold. No oscillating solution has been observed.

\begin{figure}[h]
\begin{center}
	\resizebox*{12cm}{!}{\includegraphics{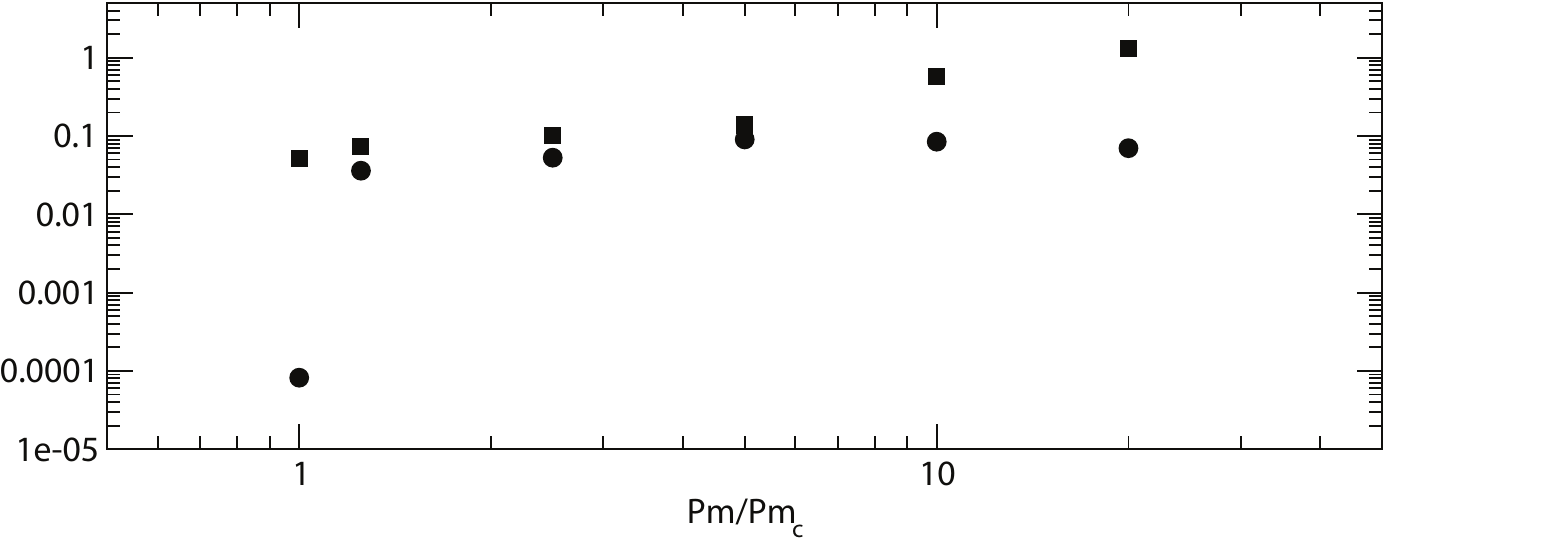}}
	\caption{Ratio of the temporally averaged values of the magnetic energy of the fluid to the kinetic energy in function of $Pm/Pm_c$ for $Re=-1500$ (circles) and $Re=-2000$ (squares) for $E=10^{-3}$.}
	\label{fig:saturE3}
\end{center}
\end{figure}

The ratio of magnetic energy to kinetic energy for several values of $Pm$ are shown in figure~\ref{fig:saturE3}. Unlike the magnetic field, the kinetic energy depends on the chosen frame. Consequently, the kinetic energies calculated in the rotating frame and in the laboratory frame are different. All the kinetic energies presented in this paper are calculated in the rotating frame. Moreover as the kinetic and magnetic energy have different scalings, a multiplicative factor $\pl 1 / \chi+1/ (Re E) \pr$ is applied on the magnetic energy to be comparable with kinetic energy. A dependence on the Rossby number can be noticed from the different behaviour of the magnetic energy for $Ro=-1.5$ and $Ro=-2$: the equipartition is never reached for $Ro=-1.5$ even for $Pm=20Pm_c$ while the magnetic energy has the same magnitude than the kinetic energy for $Ro=-2$ as soon as $Pm=10Pm_c$. The saturation of the magnetic field is complex to understand in case of global rotation as the Coriolis force does not appear in the balance of energies but is important in the balance of forces. Therefore the magnetic field may saturate before the equipartition of energy is reached for large enough magnetic Prandtl numbers. This might be especially relevant for moderate value of $|Ro|$ where the Coriolis force is predominant.

\begin{figure}[h]
\begin{center}
	\resizebox*{10cm}{!}{\includegraphics{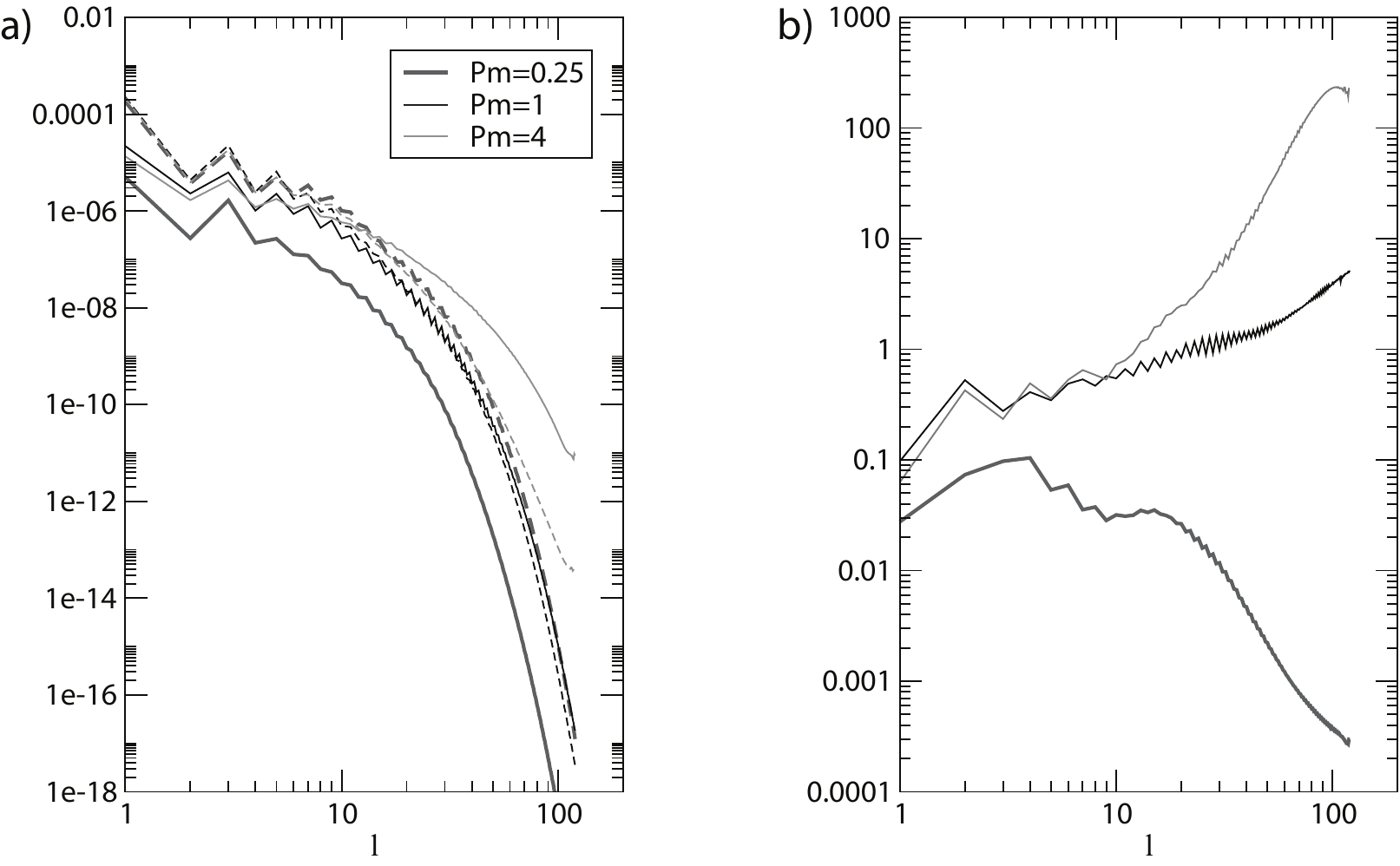}}
	\caption{(a) Magnetic (solid line) and kinetic (dashed line) energy spectra as a function of the spherical harmonic degree $l$ for the simulations at $E=10^{-3}$, $Re=-1500$ and $Pm=0.25$ (thick gray line), $Pm=1$ (black line), $Pm=4$ (thin gray line). The energies are computed over the whole shell and are time averaged. (b) Ratio of the magnetic spectrum to the kinetic one.}
	\label{fig:specE3}
\end{center}
\end{figure}

Spectra of the magnetic and kinetic energy are shown in figure~\ref{fig:specE3}. One can notice that the solutions are fully converged. The case with global rotation with a negative differential rotation ($E=10^{-3}$, $Re\sim -1500$) is the only case that allows us to study the repartition of energy at the different spatial scale in dynamos with relatively small magnetic Prandtl number ($Pm<1$) and large magnetic Prandtl number ($Pm>1$). Indeed the magnetic Prandtl number is also a measure of the kinetic dissipation over the magnetic one. Then we expect different dissipation scales depending on the magnitude of $Pm$. First we notice that the magnetic field is a large scale field as most of the magnetic energy is present in the large stuctures (spherical harmonic degree $l<10$). For the values of $Pm$ studied (from $Pm=0.25$ to $Pm=4$) the kinetic energy is always at least $2$ times greater than the magnetic field at large scale ($l<5$) (figure~\ref{fig:specE3}(b)). Both previous observations are consistent with the non equipartition of energy when increasing $Pm$ beyond the dynamo onset for these parameters ($E=10^{-3}$, $Re=-1500$, see figure~\ref{fig:saturE3}). At small $Pm$ ($Pm=0.25$), the ratio of magnetic energy to kinetic one is roughly flat from $l=1$ to $l=20$. It decreases sharply for small scale structures ($l>20$). Consequently the magnetic energy is dissipated at larger scales than the kinetic energy as expected for $Pm<1$. For $Pm=1$ the magnetic and kinetic spectra are similar for small structures ($l>20$) and their ratio is close to $1$. Finally as $Pm$ grows beyond $1$, the magnetic dissipation diminishes and small scale magnetic structures are more present. The spectra as a function of the spherical harmonic order $m$ exhibit the same behaviour. 
\newline \indent 
As for the case without global rotation, we did not find subcritical dynamos.

\subsection{Perspectives for future experimental dynamos}
\begin{figure}[h]
\begin{center}
	\resizebox*{10cm}{!}{\includegraphics{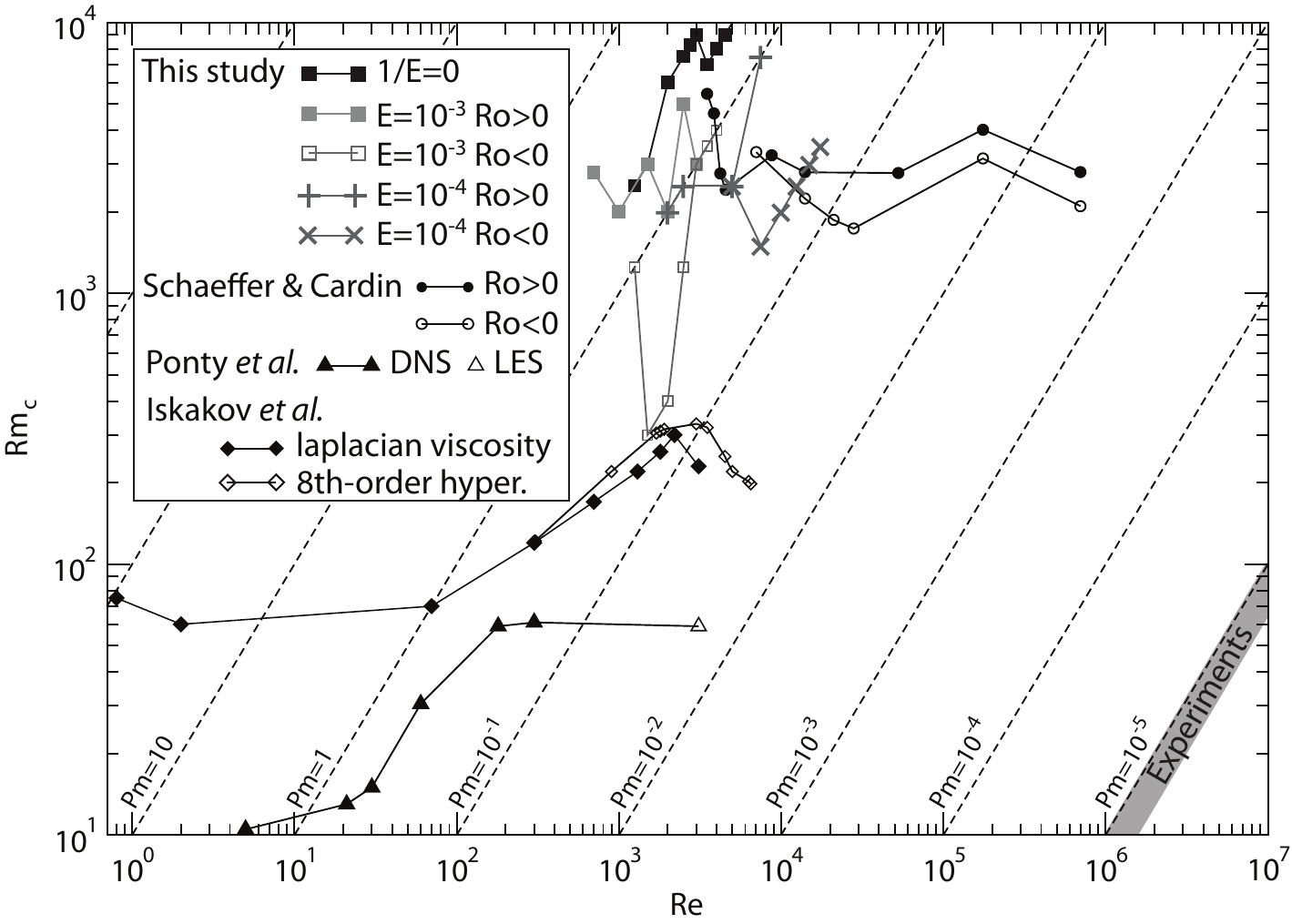}}
	\caption{Dynamos in the parameter space $Rm_c$ versus $Re$. Results from previous studies have been carried forward: \citet{Sch06} (circles), \citet{Pon07} (triangles) and \citet{Isk07} (diamonds) The highest-$Re$ point of \citet{Pon07} was obtained by large-eddy simulation. Results of \citet{Isk07} based on laplacian viscosity and 8th-order hyperviscosity are shown separately. Note that the value of $Rm$ and $Re$ for \citet{Pon07} and \citet{Isk07} are based on the rms velocity and the forcing wave number \citep[see]{Isk07}. The parameter range for experiments is roughly symbolized by the gray area.}
	\label{fig:Rm_Re}
\end{center}
\end{figure}

In experiments using liquid sodium the magnetic Prandtl number $Pm$ is of the order of $10^{-5}$ whereas in our numerical dynamo simulations $Pm$ is always larger than $0.2$. 
Decreasing the critical magnetic Prandtl number $Pm_c$ may be achieved by increasing the Reynolds number $Re$ if the critical magnetic Reynolds number $Rm_c$ remains constant. Indeed \citet{Pon07} with dynamos driven by a Taylor-Green forcing demonstrated that a dynamo threshold exists for all $Pm$ they investigated (from $1$ to $10^{-2}$)  and for high enough $Re$, $Rm_c(Re)$ remains constant (figure~\ref{fig:Rm_Re}). Using an nonhelical randomly forced turbulence, \citet{Isk07} retrieved the same result (figure~\ref{fig:Rm_Re}). In our dynamo calculations we have not found a plateau where $Rm_c(Re)$ remains constant even at $Re \sim 10^4$, which is already computationally demanding for a fully three-dimensional code.
\newline \indent
Another possibility to lower $Pm_c$ is to decrease the Ekman number $E$ as shown by \citet{Sch06}: they found dynamos at low magnetic Prandtl number (down $Pm=0.003$) by decreasing $E$ (from $10^{-6}$ to $10^{-8}$) using quasi-geostrophic flows. We can not reach such a low Ekman numbers so we worked out dynamos from $E=10^{-3}$ to $E=10^{-4}$. We were particularly interested in the evolution of the ``dynamo window'' created by the ageostrophic shear layer ($-1.5 \ge Ro \ge -2$ for $E=10^{-3}$) (figure~\ref{fig:PmRoE}) that could not be computed by Schaeffer \& Cardin with a QG model. As previously discussed, dynamos beneath $Pm=0.2$ have not been found and the magnetic Reynolds number needed for $E=10^{-4}$ is higher than for $E=10^{-3}$. We interpret the disapperance (for $E=10^{-4}$) of the dynamo window (present at $E=10^{-3}$) by the destabilization of the shear layer because of the emergence of small scale fluctuations due the higher $Re$ in a regime where the Proudman-Taylor constraint fails ($|Ro|\ge 1$). 
Nevertheless decreasing $E$ lowers $Pm_c$ in the positive differential rotation case (from $Pm_c \sim 1$ for $E=10^{-3}$ to $Pm_c \sim 0.4$ for $E=10^{-4}$). The simulations presented in this paper require high resolution and the current numerical limits do not allow us to carry on this study towards more rapidly rotating system or more boundary forcing. New phenomena may arise for $E \ll 1$ or $Re \gg Re_c$. Without evidence, we may speculate that the results obtained with a QG approach \citep{Sch06} are the extension of our work towards small $E$. In that hypothesis a dynamo branch of constant $Rm_c$ can be reached for small Ekman numbers ($E<10^{-6}$) and for moderate Rossby numbers ($|Ro|<0.1$). Moreover the presence of the peak of induced magnetic field when the fluid rotation rate cancels the rotation rate of the outer sphere in the DTS experiment \citep{Nat08} suggests that new ``dynamo windows'' may exist for small $E$ ($E\sim 10^{-7}$).
To conclude, our results tend to show that dynamos at low magnetic Prandtl number may be achieved when decreasing the Ekman number but it likely requires high magnetic Reynolds number (a few thousand) which is far from what could be obtained in liquid sodium experiments (upper limit of $Rm$ of the order of $700$ for the new $3$-m diameter experiment of the College Park group in Maryland).

\section{Ferromagnetic boundary conditions}
\label{sec:ferro} 
Our interest in studying the influence of ferromagnetic boundary conditions on the dynamo onset stems from the recent success of the Von K\'arm\'an Sodium (VKS) experiment \citep{Mon07} in producing a self-sustained magnetic field. The use of soft iron (ferromagnetic) propellers instead of stainless steel (amagnetic) ones of the same geometry facilitates the generation of a magnetic field. A ferromagnetic medium has a magnetic permeability larger than the vaccum permeability. When increasing the magnetic permeability of a boundary the magnetic field lines in the fluid tend to become perpendicular to this boundary. Numerical simulations \citep{Gis08,Lag08} with impellers of infinite magnetic permeability i.e. imposing a purely normal magnetic field at the boundary have shown that the critical magnetic Reynolds number for dynamo action is lowered due to channeling of the magnetic field lines in the region of the strongest shear. Moreover \citet{Gis08} predicted that adding ferromagnetic walls would also lower the threshold significatively. Similarly, \citet{Ava03} found that the addition of a ferromagnetic wall in a Ponomarenko dynamo leads to a reduction of the dynamo onset. However, in Morin's simulations of convective dynamos in a spherical shell \citep{Mor05}, the effect of ferromagnetic boundaries is more complicated and there are several possibilities: a purely radial magnetic field at the outer boundary seems to favour dynamo action whereas imposing a radial field on both inner and outer boundaries inhibits dynamo action. Therefore no general conclusions can be drawn about the impact of high magnetic permeability boundary conditions. We may conjecture that if ferromagnetic pieces concentrate magnetic field lines away from the region where dynamo processes take place, their presence is useless. Their local action of concentration of magnetic field lines has to be roughly coincident with induction process in the liquid.
\newline \indent
We study different cases for their relevance to experimental setup. The ferromagnetic core with an insulating outer sphere case is investigated for its simplicity to put into pratice in experiments. Moreover we expect that the impact of a change of the magnetic boundary conditions will be greater if the boundary is close to the place where the magnetic field is generated i.e. near the inner core where the angular shear is maximum. The ferromagnetic outer sphere with a non-ferromagnetic conducting core is also studied since the contact surface is larger. To complete the study, the case with both inner and outer ferromagnetic boundaries is also investigated. Note that in these simulations, the conductivity of any ferromagnetic media is set equal to that of the fluid, so we can study the effect of the ferromagnetic boundaries in isolation.
\newline \indent
Suppose there is a jump in the magnetic permeability at a radius $r$, representing the interface between two media denoted $1$ and $2$ of magnetic permeability $\mu_1$ and $\mu_2$ respectively. We then have the following boundary equations for the poloidal $B_p$ and toroidal $B_t$ scalars
\begin{eqnarray}
  B_p^1(r)&=&B_p^2(r),
\\
  B_t^1(r)&=&\frac{\mu_1}{\mu_2} B_t^2(r),
\label{eq:ferro_Btor}
\\
  \left. \frac{1}{r}\frac{\partial \left(r B_p^1\right)}{\partial r}\right|_r 
    &=& \frac{\mu_1}{\mu_2}\left. \frac{1}{r}\frac{\partial \left(r B_p^2\right)}{\partial r}\right|_r,
\end{eqnarray}
where the superscripts $1$ and $2$ stand for the regions. The discontinuity of the toroidal and spheroidal scalars is carefully implemented in the time evolution scheme and in the non-linear terms.
\newline \indent
The reader may note that we do not include a remanent magnetic field in our model, although such a field can exist within ferromagnetic material. In order to assess the impact of a remanent magnetic field, we carried out simulations with an imposed axial dipolar magnetic field in the inner core separately from the problem of the ferromagnetic boundary conditions. We found that the presence of the imposed magnetic field have no effect on the dynamo onset in this geometry.
\newline \indent
We focus our study on the dynamo threshold; thus the results presented are only for simulations near the dynamo onset.

\subsection{Ferromagnetic inner core}
\begin{figure}
\begin{center}
	\resizebox*{12cm}{!}{\includegraphics{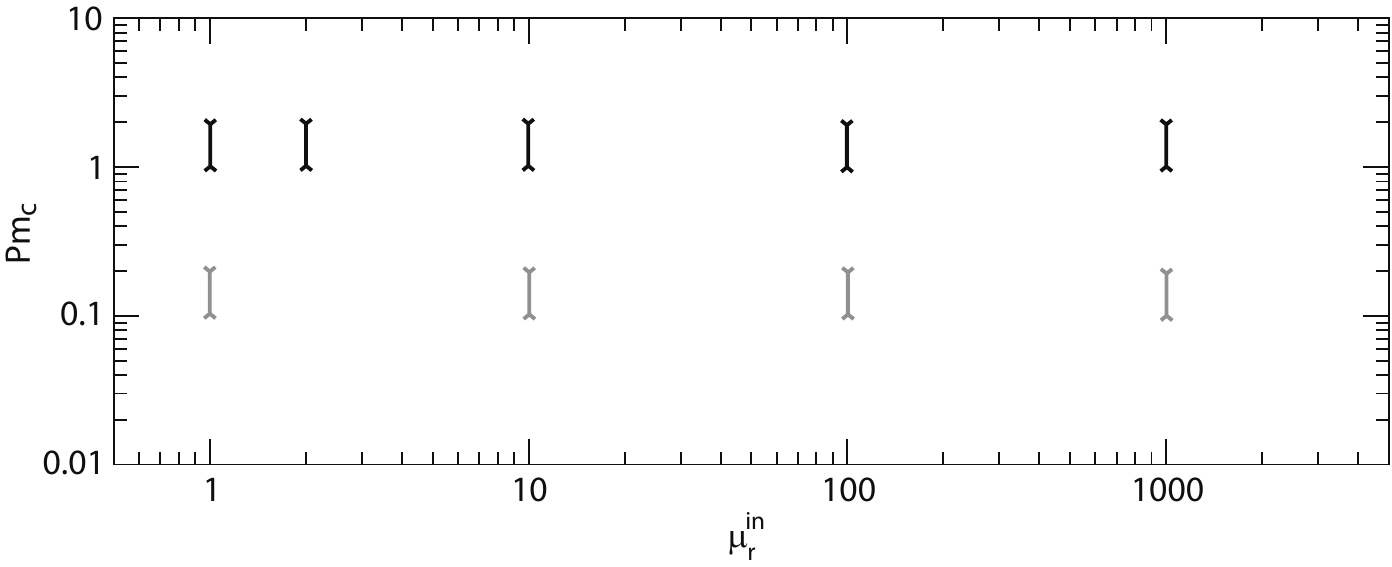}}
	\caption{Critical Prandlt number $Pm_c$ in function of the relative magnetic permeability of the inner core $\mu_r^{in}$ for a case without global rotation ($Re=1500$) (black) and with global rotation ($E=10^{-3}$, $Re=-1500$) (gray). The bottom of the bar corresponds to a non dynamo calculation whereas the top corresponds to a dynamo calculation.}
	\label{fig:threshold_in}
\end{center}
\end{figure}

The core is assumed to have the magnetic permeability $\mu_{in}$, so that the relative magnetic permeability is $\mu_r^{in}=\mu_{in}/\mu_0$. The core has still the same conductivity as the fluid, and the outer sphere is insulating and non-ferromagnetic. Note that the magnetic permeability of soft iron is about $1000\mu_0$.
\newline \indent 
The main result is shown in figure~\ref{fig:threshold_in}: whatever are the relative magnetic permeability ($\mu_r^{in}=1$ to $\mu_r^{in}=1000$), the dynamo onsets are very similar to the ones computed without ferromagnetic parts. Note that the results presented in these section have been verified with or without global rotation (simulations at $E=10^{-3}$ with positive or negative differential rotation of the inner core). Nevertheless for simulations just below the dynamo onset the decay rate is decreased when the relative magnetic permeability $\mu_r^{in}$ is increased (not shown). The critical magnetic Reynolds numbers involved in these dynamos are quite high so we only determine them in a coarse range. A marginal effect of ferromagnetic boundaries may be more important than what we report but, in this paper, we are only interested in significant changes.

\begin{figure}
\begin{center}
	\resizebox*{12cm}{!}{\includegraphics{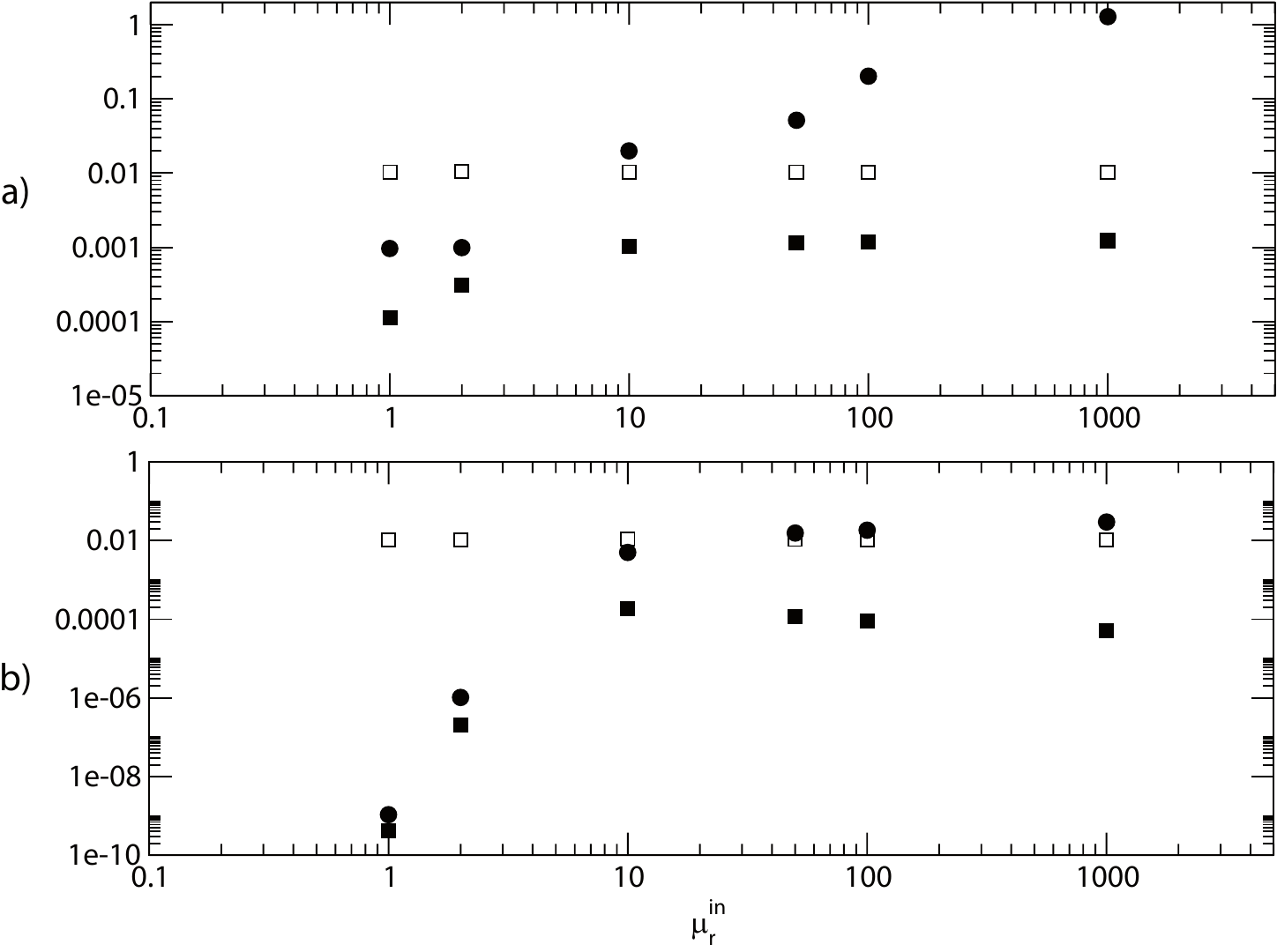}}
	\caption{Kinetic energy (open squares) and magnetic energy in the fluid (filled squares) and in the inner core (filled circles) in function of the magnetic permeability of the core $\mu_r^{in}$ for the parameters: (a) $1/E=0$, $Re=1500$, $Pm=2$ and (b) $E=10^{-3}$, $Re=-1500$ and $Pm=0.2$. The energies are dimensionless; the scaling factor is $\rho(r_i \Delta \Omega)^2$.}
	\label{fig:mu_in}
\end{center}
\end{figure}

The kinetic energy remains roughly constant when $\mu_r^{in}$ is increased (figures~\ref{fig:mu_in}). Without overall rotation (figure~\ref{fig:mu_in}(a)) the magnetic energy of the fluid increases by a factor $10$ when $\mu_r^{in}$ rises from $1$ to $10$, and then remains constant (about 10 times smaller than the kinetic energy) as $\mu_r^{in}$ is increased further. For $\mu_r^{in}\le 2$ the dynamo field is an oscillating quadrupole as described in section~\ref{sec:magE0_a}. For $\mu_r^{in}\ge 10$ the oscillating behaviour of the magnetic field is lost, and the dynamo field is a stationary axial quadrupole. Consequently, the oscillating quadrupolar dynamo is restricted to the weak magnetic field regime as we report in section~\ref{sec:magE0_a} when observing the effect of $Pm$ on energy partition.
With global rotation (figure~\ref{fig:mu_in}(b)) the effect of a ferromagnetic inner core is similar in positive (not shown) and negative differential rotation: after an increase of the magnetic energy when $\mu_r^{in}$ increases from 1 to 10, the magnetic energy of the fluid decreases slightly when $\mu_r^{in}>10$. The poloidal and toroidal magnetic energies in the fluid follow the same evolution but the toroidal energy is always around $3$ times greater than the poloidal one (not shown).
\newline \indent 
The magnetic energy in the inner core raises as $\mu_r^{in}$ increases and is $500$ to $1000$ times greater in the core than in the fluid for $\mu_r^{in}=1000$. Note that the values presented here take into account the variation of magnetic permeability of the inner core $\mu_{in}$. 

\begin{figure}
\begin{center}
	\resizebox*{10cm}{!}{\includegraphics{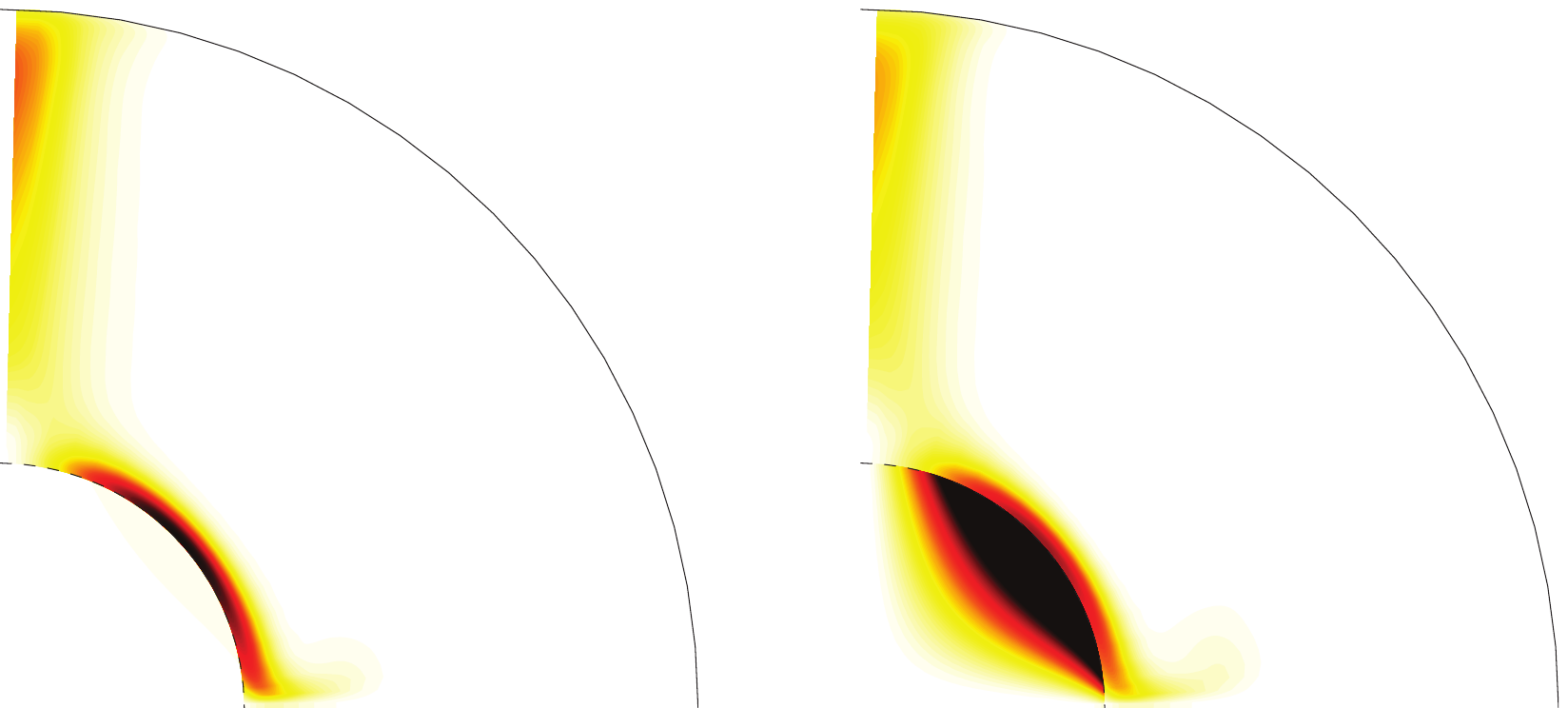}}
	\caption{Magnetic field zonally averaged in a quadrant of the meridional plane for $E=10^{-3}$, $Re=1000$, $Pm=2$, $\mu_r^{in}=1$ (left) and $\mu_r^{in}=10$ (right). The color scale is the same for both figures. Dark colors indicate a strong magnetic field.}
	\label{fig:ME_in}
\end{center}
\end{figure}

The little influence of the change of the magnetic boundary conditions on the dynamo generation can be understood by the interpretation of figure~\ref{fig:ME_in}. Indeed we observe that increasing the magnetic permeability of the inner core does not change significatively the morphology of the axisymmetric magnetic field in the fluid. However as the toroidal field in the fluid is barely changed, the toroidal field inside the inner core is increased by a factor $\mu_r^{in}$ (equation (\ref{eq:ferro_Btor})). We notice that the mechanical work exerted by the inner core is increased when raising the magnetic permeability. As it is equal to the total dissipation in the statistically saturated regime, we conclude that the ohmic dissipation inside the inner core is enhanced. This is consistent with the presence of a large magnetic energy in the inner core.

\begin{figure}
\begin{center}
	\resizebox*{10cm}{!}{\includegraphics{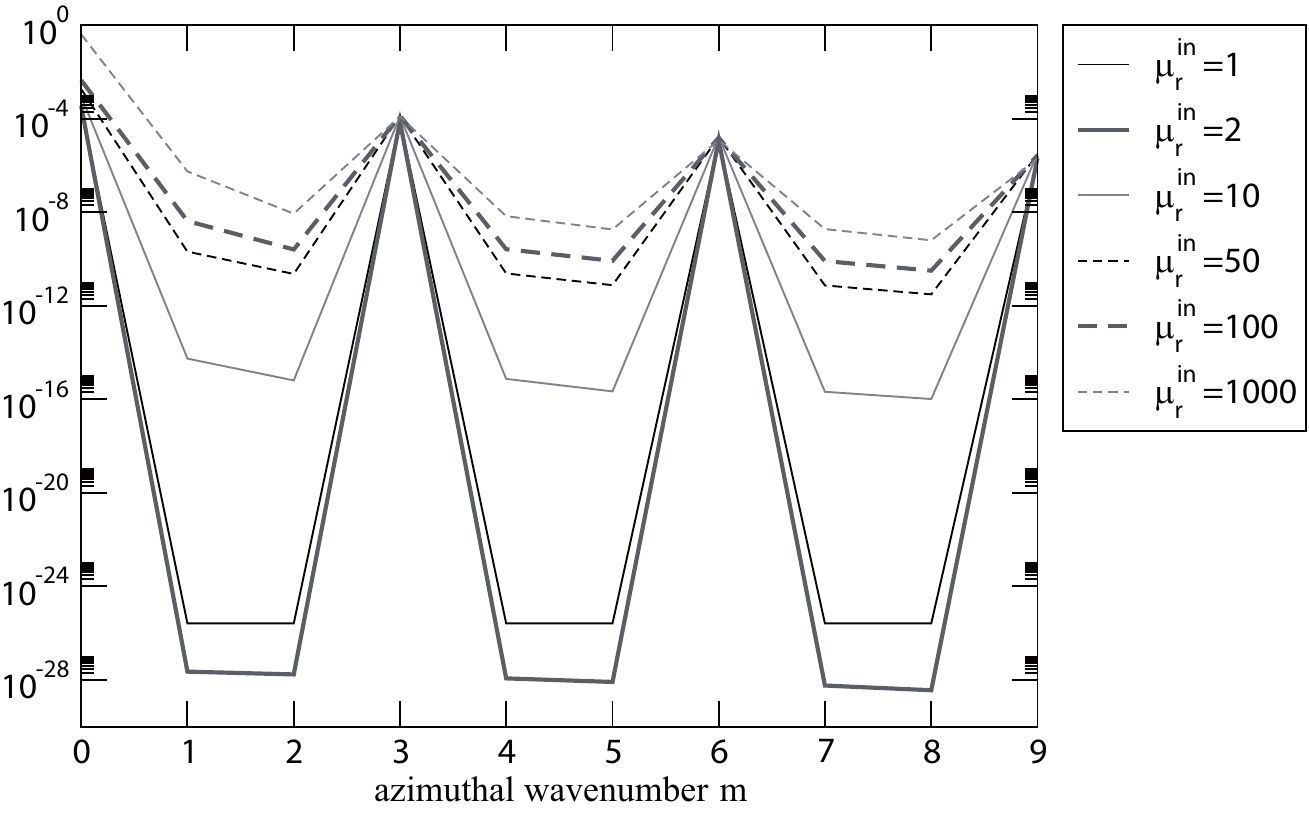}}
	\caption{Magnetic energy spectra in function of the azimuthal wavenumber $m$ for simulations at $1/E=0$, $Re=1500$, $Pm=2$ and for different relative magnetic permeability $\mu_r^{in}$. The power spectra are radially averaged on the fluid shell.}
	\label{fig:spec_in}
\end{center}
\end{figure}

The magnetic energy spectra (figure~\ref{fig:spec_in}) show that the magnetic field which is initially ($\mu_r^{in}=1$) partly dominated by the most unstable kinetic mode ($m=3$ for $Re=1500$ and $1/E=0$) becomes strongly axisymmetric when increasing the magnetic permeability $\mu_r^{in}$. This behaviour is due to the fact that magnetic field lines are maintained by the ferromagnetic core. They are expelled in the polar area and then concentrated close to the rotation axis.

\subsection{Ferromagnetic outer sphere}
\begin{figure}
\begin{center}
	\resizebox*{12cm}{!}{\includegraphics{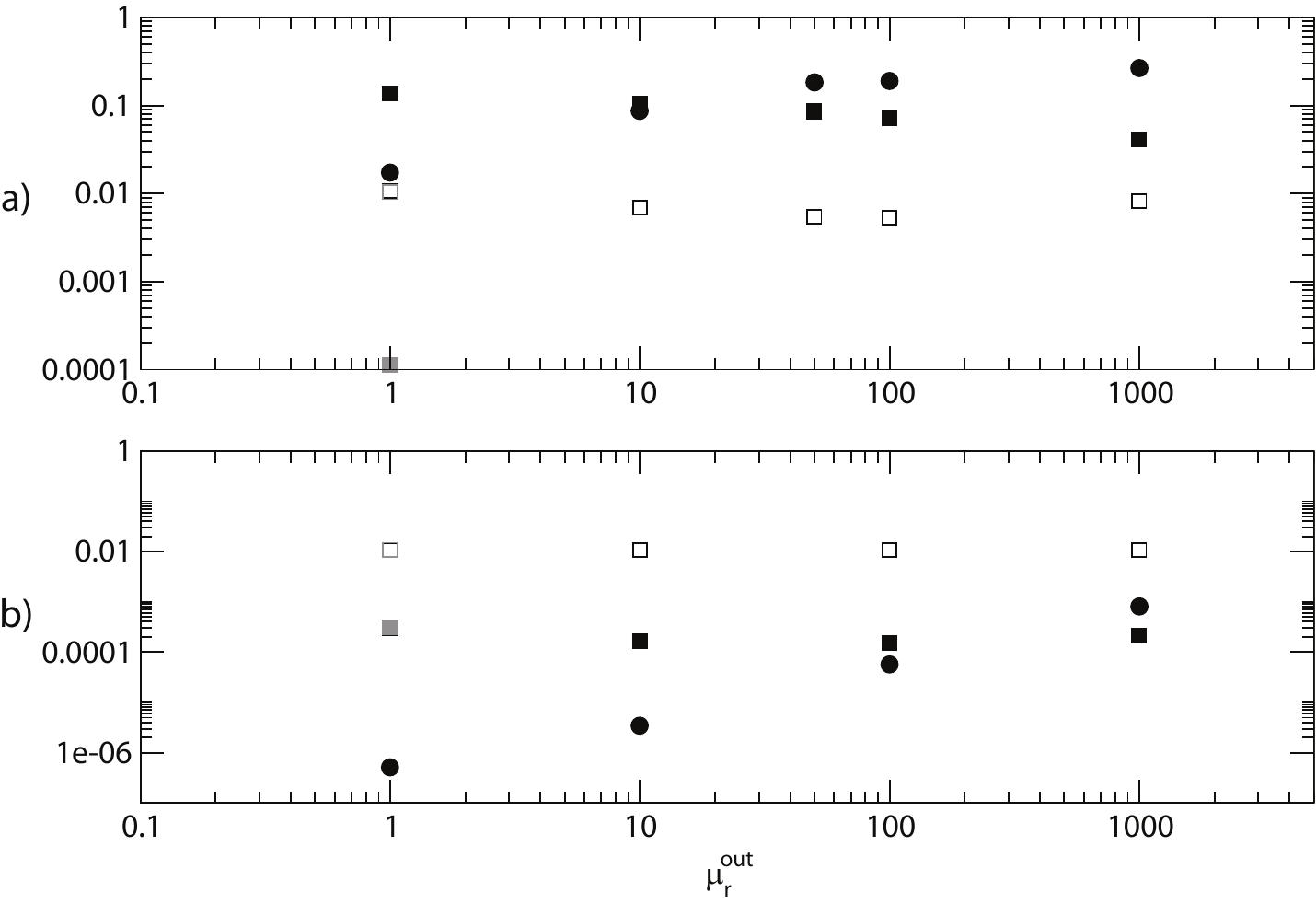}}
	\caption{Kinetic energy (open black squares) and magnetic energy in the fluid (filled black squares) and in the outer shell (filled black circles) as a function of the relative magnetic permeability of the outer shell $\mu_r^{out}$ for the parameters: (a) $1/E=0$, $Re=1500$, $Pm=2$ and (b) $E=10^{-3}$, $Re=-1500$, $Pm=0.25$. The gray symbols stand for the kinetic energy (open square) and the magnetic energy (filled square) when the outer shell is non-ferromagnetic and non-conducting. They cover the black symbols for $\mu_r^{out}=1$ in plot b. The energies are dimensionless; the scaling factor is $\rho(r_i \Delta \Omega)^2$.}
	\label{fig:mu_out}
\end{center}
\end{figure}

The outer sphere is assumed to have the same conductivity as the fluid and a relative magnetic permeability $\mu_r^{out}=\mu_{out}/\mu_0$ where $\mu_{out}$ is the magnetic permeability of the outer sphere. The thickness of the shell is arbitrarily fixed at $0.08r_o$. Simulations with an outer sphere thickness of $0.02r_o$ produce similar results. The magnetic outer boundary conditions match a potential field outside in the vaccum. The inner core has the same electric conductivity and magnetic permeability $\mu_0$ as the fluid. In order to assess the impact of a conducting outer sphere alone we have carried out simulations with $\mu_r^{out}=1$. 
\newline \indent
With or without global rotation the threshold remains unchanged when imposing a relative magnetic permeability $\mu_r^{out}$ up to $1000$ with both positive and negative differential rotation.
\newline \indent
When a global rotation is imposed the magnetic energy in the fluid is constant (figure~\ref{fig:mu_out}(b)) whereas the magnetic energy in the outer sphere increases by a factor $1000$ when $\mu_r^{out}=1000$. Adding a conductive outer sphere has no noticeable impact on the threshold: both magnetic and kinetic energies are the same as in the case with a non-conducting and non-ferromagnetic outer sphere. The morpholgy of the magnetic field is very similar to the case with the ferromagnetic inner core with the dominance of the axisymmetric component of the field when the relative magnetic permeability is increased and magnetic field lines concentrated around the rotation axis.
\newline \indent
Without overall rotation, the magnetic energy lies at a high value, $5$ to $15$ times greater than the kinetic energy with smaller values when $\mu_r^{out} \ge 100$ (figure~\ref{fig:mu_out}(a)). The presence of a conducting outer sphere yields an induced magnetic field $1000$ times greater than without a conducting outer sphere. 

\subsection{Inner and outer ferromagnetic boundary conditions}
\begin{figure}
\begin{center}
	\resizebox*{12cm}{!}{\includegraphics{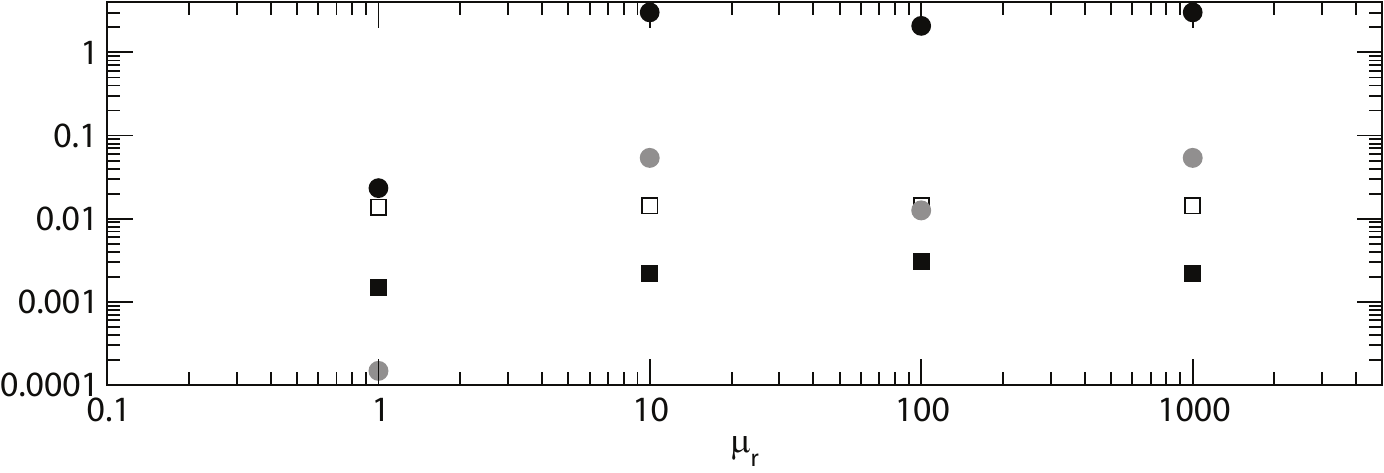}}
	\caption{Kinetic energy (open black squares) and magnetic energy in the fluid (filled black squares), in the inner core (filled black circles) and in the outer shell (filled gray circles) as a function of the relative magnetic permeability of both inner and outer boundaries $\mu_r$ for the parameters $E=10^{-3}$, $Re=1000$ and $Pm=2$. The energies are dimensionless; the scaling factor is $\rho(r_i \Delta \Omega)^2$.}
	\label{fig:E3mu_both}
\end{center}
\end{figure}

In this last case, both outer and inner boundaries are ferromagnetic (with same magnetic permeability) and conducting (with the same conductivity as the fluid). 
\newline \indent
As previously the dynamo onset is unchanged when the relative magnetic permeability is increased with or without global rotation. The kinetic and magnetic energy in the fluid remains constant (figure~\ref{fig:E3mu_both}). The magnetic energy in the inner core and in the outer sphere increases by a factor $100$ when the relative magnetic permeability $\mu_r$ is raised from $1$ to $10$ and then remains roughly constant when $\mu_r$ is further increased.

\section{Conclusions}
Using laboratory experiments to model planetary cores is difficult; modelling the natural forcing of the planetary cores such as thermal and chemical convection, requires high power in laboratory as the efficiency to convert the power into kinetic energy is low in convective systems \cite[see for instance][]{Aub01,Gil06}. Consequently, mechanical forcings are necessary and spherical Couette flows are suitable to answer key questions about spherical natural dynamos. Without global rotation, our numerical study shows that the dynamo action is always found for high magnetic Reynolds number ($Rm_c$ greater than a few thousand) for Reynolds numbers $Re>Re_c$ at which the first non-axisymmetric hydrodynamical instability develops. The increase of the boundary forcing (measured by the Reynolds number) barely changes the critical magnetic Prandtl number of the Couette dynamo. In other words, the critical magnetic Reynolds number increases with the Reynolds number of the Couette flow according to our calculations. The overall rotation of the spherical Couette system yields a slight decrease of the dynamo threshold except for a regime where relatively low magnetic Reynolds numbers ($Rm\sim 300$) have been found. For moderate Ekman number ($E=10^{-3}$) this dynamical regime corresponds to a state where the geostrophic constraint fails, i.e. the inertial terms associated to a negative differential rotation balance the Coriolis force associated to the positive overall rotation ($Ro\sim -1$). In this regime, large shear in the vicinity of the inner core is present and may explain the large decrease of the dynamo onset. For smaller Ekman number ($E=10^{-4}$), a strong boundary forcing is necessary to reach the state $Ro\sim-1$. As the geostrophic constraint is lost, the enhanced shear at the inner core is unstable and generates small scale 3D fluctuations which do not improve the self generation of the magnetic field.
\newline \indent
Without global rotation, the equipartition between magnetic and kinetic energy is achieved either for $Re \gtrsim 2Re_c$ or for $Pm \gtrsim 2Pm_c$. The interaction parameter (ratio of the Lorentz forces to inertia) is of the order of unity and the Lorentz forces modify the velocity field. Spherical Couette flows are thus able to generate a strong magnetic field when no global rotation is imposed. This observation is of first order interest for experiments, as in planetary dynamos the Lorentz forces are one of the leading forces in the system. In the previous and current dynamo experiments, the interaction parameter is much smaller than unity (around $10^{-4}$ for instance in the Von K\'arm\'an Sodium dynamo experiment \citep{Mon07}). With global rotation, the presence of the Coriolis force might lead to the saturation of the magnetic field before the equipartition of energy. Accordingly the interaction parameter is smaller than unity unless the hydrodynamical regime is far from the regime where the Coriolis force is predominant.
\newline \indent
Despite previous studies mostly highlighting the positive impact of ferromagnetic media in the setup for the generation of the magnetic field, our numerical results demonstrate that adding high magnetic permeability boundary conditions have no impact on the threshold in all the configurations studied. Nevertheless the amplitude of the generated magnetic field is enhanced, a feature that can be useful for experiments in order to obtain a stronger dynamo magnetic field. But producing a stronger induced magnetic field thanks to the use of ferromagnetic pieces in an experiment has also a cost: the increase of the ohmic dissipation in the experiment implies an increase of the power needed to run the experiment which could be problematic.
\vspace{6pt}

\section*{Acknowledgements}
We thank the geodynamo group in Grenoble for useful discussions. We are grateful for the comments of the reviewers, which have greatly improved the quality of this paper. All the computations presented in this paper were performed at the Service Commun de Calcul Intensif de l'Observatoire de Grenoble (SCCI). Financial support was provided by CNRS/INSU. CG was supported by a research studentship from Universit\'e Joseph-Fourier. 
This is a preprint of an article whose final and definitive form has been published in Geophysical and Astrophysical fluid dynamics \copyright 2010 \copyright Taylor \& Francis. Geophysical and Astrophysical fluid dynamics is available online at: \href{http://www.informaworld.com/smpp/}{http://www.informaworld.com/smpp/}.

\bibliographystyle{plain}
\bibliography{ref}

\begin{thebibliography}{43}
\providecommand{\natexlab}[1]{#1}

\bibitem[\protect\citeauthoryear{{Aubert} {\itshape{et~al.}}}{2001}]{Aub01}
{Aubert}, J., {Brito}, D., {Nataf}, H.C., {Cardin}, P. and {Masson}, J.P., {A
  systematic experimental study of rapidly rotating spherical convection in
  water and liquid gallium}. {\itshape Phys. Earth and Planet. Int.} 2001,
  \textbf{128}, 51--74.

\bibitem[\protect\citeauthoryear{{Avalos-Zuniga}
  {\itshape{et~al.}}}{2003}]{Ava03}
{Avalos-Zuniga}, R., {Plunian}, F. and {Gailitis}, A., {Influence of
  electromagnetic boundary conditions onto the onset of dynamo action in
  laboratory experiments}. {\itshape Phys. Rev. E} 2003, \textbf{68}, 066307.

\bibitem[\protect\citeauthoryear{{Bourgoin} {\itshape{et~al.}}}{2002}]{Bou02}
{Bourgoin}, M., {Mari{\'e}}, L., {P{\'e}tr{\'e}lis}, F., {Gasquet}, C.,
  {Guigon}, A., {Luciani}, J.B., {Moulin}, M., {Namer}, F., {Burguete}, J.,
  {Chiffaudel}, A., {Daviaud}, F., {Fauve}, S., {Odier}, P. and {Pinton}, J.F.,
  {Magnetohydrodynamics measurements in the von K{\'a}rm{\'a}n sodium
  experiment}. {\itshape Phys. Fluids} 2002, \textbf{14}, 3046--3058.

\bibitem[\protect\citeauthoryear{{Brito} {\itshape{et~al.}}}{2001}]{Bri01}
{Brito}, D., {Nataf}, H.C., {Cardin}, P., {Aubert}, J. and {Masson}, J.P.,
  {Ultrasonic Doppler velocimetry in liquid gallium}. {\itshape Exp. Fluids}
  2001, \textbf{31}, 653--663.

\bibitem[\protect\citeauthoryear{Christensen and Aubert}{2006}]{Chr06}
Christensen, U.R. and Aubert, J., Scaling properties of convection-driven
  dynamos in rotating spherical shells and application to planetary magnetic
  fields. {\itshape Geophy. J. Int.} 2006, \textbf{166}, 97--114.

\bibitem[\protect\citeauthoryear{{Christensen}
  {\itshape{et~al.}}}{2001}]{Chr01}
{Christensen}, U.R., {Aubert}, J., {Cardin}, P., {Dormy}, E., {Gibbons}, S.,
  {Glatzmaier}, G.A., {Grote}, E., {Honkura}, Y., {Jones}, C., {Kono}, M.,
  {Matsushima}, M., {Sakuraba}, A., {Takahashi}, F., {Tilgner}, A., {Wicht}, J.
  and {Zhang}, K., {A numerical dynamo benchmark}. {\itshape Phys. Earth
  Planet. Inter.} 2001, \textbf{128}, 25--34.

\bibitem[\protect\citeauthoryear{{Dormy}}{1997}]{Dor97}
{Dormy}, E., {Mod\'elisation num\'erique de la dynamo terrestre}. PhD thesis,
  Institut de Physique du Globe de Paris 1997.

\bibitem[\protect\citeauthoryear{{Dormy} {\itshape{et~al.}}}{1998}]{Dor98}
{Dormy}, E., {Cardin}, P. and {Jault}, D., {MHD flow in a slightly
  differentially rotating spherical shell, with conducting inner core, in a
  dipolar magnetic field}. {\itshape Earth Planet. Sci. Lett.} 1998,
  \textbf{160}, 15--30.

\bibitem[\protect\citeauthoryear{{Dudley} and {James}}{1989}]{Dud89}
{Dudley}, M.L. and {James}, R.W., {Time-dependent kinematic dynamos with
  stationary flows}. {\itshape Proc. R. Soc. Lond. A} 1989, \textbf{425},
  407--429.

\bibitem[\protect\citeauthoryear{{Dumas}}{1991}]{Dum91}
{Dumas}, G., {Study of Spherical Couette Flow via Three-Dimensional Spectral
  Simulations: Large and Narrow-Gap Flows and Their Transitions}. PhD thesis,
  California Institute of Technology 1991.

\bibitem[\protect\citeauthoryear{{Egbers} and {Rath}}{1995}]{Egb95}
{Egbers}, C. and {Rath}, H.J., {The existence of Taylor vortices and wide-gap
  instabilities in spherical Couette flow}. {\itshape Acta Mech.} 1995,
  \textbf{111}, 125--140.

\bibitem[\protect\citeauthoryear{{Forest} {\itshape{et~al.}}}{2002}]{For02}
{Forest}, C.B., {Bayliss}, R.A., {Kendrick}, R.D., {Nornberg}, M.D.,
  {O'Connell}, R. and {Spence}, E.J., {Hydrodynamic and numerical modeling of a
  spherical homogeneous dynamo experiment}. {\itshape Magnetohydrodynamics}
  2002, \textbf{38}, 107--120.

\bibitem[\protect\citeauthoryear{{Gagni\`ere}}{2009}]{Gag09}
{Gagni\`ere}, N., {Etude exp\'erimentale du r\'egime magn\'etostrophique avec
  DTS (Derviche Tourneur Sodium)}. PhD thesis, Universit\'e Joseph-Fourier -
  Grenoble 2009.

\bibitem[\protect\citeauthoryear{{Gailitis} {\itshape{et~al.}}}{2001}]{Gai01}
{Gailitis}, A., {Lielausis}, O., {Platacis}, E., {Dement'ev}, S., {Cifersons},
  A., {Gerbeth}, G., {Gundrum}, T., {Stefani}, F., {Christen}, M. and {Will},
  G., {Magnetic field saturation in the Riga dynamo experiment}. {\itshape
  Phys. Rev. Lett.} 2001, \textbf{86}, 3024--3027.

\bibitem[\protect\citeauthoryear{{Gillet} and {Jones}}{2006}]{Gil06}
{Gillet}, N. and {Jones}, C.A., {The quasi-geostrophic model for rapidly
  rotating spherical convection outside the tangent cylinder}. {\itshape J.
  Fluid Mech.} 2006, \textbf{554}, 343--369.

\bibitem[\protect\citeauthoryear{{Gissinger} {\itshape{et~al.}}}{2008}]{Gis08}
{Gissinger}, C., {Iskakov}, A., {Fauve}, S. and {Dormy}, E., {Effect of
  magnetic boundary conditions on the dynamo threshold of von K{\'a}rm{\'a}n
  swirling flows}. {\itshape Europhys. Lett.} 2008, \textbf{82}, 29001.

\bibitem[\protect\citeauthoryear{{Heimpel} {\itshape{et~al.}}}{2005}]{Hei05}
{Heimpel}, M., {Aurnou}, J. and {Wicht}, J., {Simulation of equatorial and
  high-latitude jets on Jupiter in a deep convection model}. {\itshape Nature}
  2005, \textbf{438}, 193--196.

\bibitem[\protect\citeauthoryear{{Hide} and {Titman}}{1967}]{Hid67}
{Hide}, R. and {Titman}, C.W., {Detached shear layers in a rotating fluid}.
  {\itshape J. Fluid Mech.} 1967, \textbf{29}, 39--60.

\bibitem[\protect\citeauthoryear{{Hollerbach}}{2003}]{Hol03}
{Hollerbach}, R., {Instabilities of the Stewartson layer Part 1. The dependence
  on the sign of Ro}. {\itshape J. Fluid Mech.} 2003, \textbf{492}, 289--302.

\bibitem[\protect\citeauthoryear{{Hollerbach} {\itshape{et~al.}}}{2007}]{Hol07}
{Hollerbach}, R., {Canet}, E. and {Fournier}, A., {Spherical Couette flow in a
  dipolar magnetic field}. {\itshape Europ. J. Mech. B} 2007.

\bibitem[\protect\citeauthoryear{{Hollerbach} {\itshape{et~al.}}}{2004}]{Hol04}
{Hollerbach}, R., {Futterer}, B., {More}, T. and {Egbers}, C., {Instabilities
  of the Stewartson layer Part 2. Supercritical mode transitions}. {\itshape
  Theor. Comput. Fluid Dyn.} 2004, \textbf{18}, 197--204.

\bibitem[\protect\citeauthoryear{{Hollerbach} {\itshape{et~al.}}}{2006}]{Hol06}
{Hollerbach}, R., {Junk}, M. and {Egbers}, C., {Non-axisymmetric instabilities
  in basic state spherical Couette flow}. {\itshape Fluid Dyn. Res.} 2006,
  \textbf{38}, 257--273.

\bibitem[\protect\citeauthoryear{{Hollerbach} and {R{\"u}diger}}{2005}]{Hol05}
{Hollerbach}, R. and {R{\"u}diger}, G., {New Type of Magnetorotational
  Instability in Cylindrical Taylor-Couette Flow}. {\itshape Phys. Rev. Lett.}
  2005, \textbf{95}, 124501.

\bibitem[\protect\citeauthoryear{{Iskakov} {\itshape{et~al.}}}{2007}]{Isk07}
{Iskakov}, A.B., {Schekochihin}, A.A., {Cowley}, S.C., {McWilliams}, J.C. and
  {Proctor}, M.R.E., {Numerical Demonstration of Fluctuation Dynamo at Low
  Magnetic Prandtl Numbers}. {\itshape Phys. Rev. Lett.} 2007, \textbf{98},
  208501.

\bibitem[\protect\citeauthoryear{{Kelley} {\itshape{et~al.}}}{2007}]{Kel07}
{Kelley}, D., {Triana}, S.A., {Zimmerman}, D., {Tilgner}, A. and {Lathrop}, D.,
  {Inertial waves driven by differential rotation in a planetary geometry}.
  {\itshape Geophys. Astrophys. Fluid Dyn.} 2007, \textbf{101}, 469--487.

\bibitem[\protect\citeauthoryear{{Khlebutin}}{1968}]{Khl68}
{Khlebutin}, G.N., {Stability of fluid motion between a rotating and a
  stationnary concentric sphere}. {\itshape Fluid Dyn.} 1968, \textbf{3},
  31--32.

\bibitem[\protect\citeauthoryear{{Laguerre} {\itshape{et~al.}}}{2008}]{Lag08}
{Laguerre}, R., {Nore}, C., {Ribeiro}, A., {L{\'e}orat}, J., {Guermond}, J.L.
  and {Plunian}, F., {Impact of Impellers on the Axisymmetric Magnetic Mode in
  the VKS2 Dynamo Experiment}. {\itshape Phys. Rev. Lett.} 2008, \textbf{101},
  104501.

\bibitem[\protect\citeauthoryear{{Laure} {\itshape{et~al.}}}{2000}]{Lau00}
{Laure}, P., {Chossat}, P. and {Daviaud}, F., {Generation of magnetic field in
  the Couette–Taylor system}; in {\itshape Dynamo and dynamics, a
  mathematical challenge}, edited by P.~{Chossat}, D.~{Armbruster} and
  I.~{Oprea}, Vol. ~26 of {\itshape Nato Sci Ser. II} 2000, pp. 17--24.

\bibitem[\protect\citeauthoryear{{Le Mou{\"e}l}}{1984}]{LeM84}
{Le Mou{\"e}l}, J.L., {Outer-core geostrophic flow and secular variation of
  Earth's geomagnetic field}. {\itshape Nature} 1984, \textbf{311}, 734--735.

\bibitem[\protect\citeauthoryear{{Monchaux} {\itshape{et~al.}}}{2007}]{Mon07}
{Monchaux}, R., {Berhanu}, M., {Bourgoin}, M., {Moulin}, M., {Odier}, P.,
  {Pinton}, J.F., {Volk}, R., {Fauve}, S., {Mordant}, N., {P{\'e}tr{\'e}lis},
  F., {Chiffaudel}, A., {Daviaud}, F., {Dubrulle}, B., {Gasquet}, C.,
  {Mari{\'e}}, L. and {Ravelet}, F., {Generation of a Magnetic Field by Dynamo
  Action in a Turbulent Flow of Liquid Sodium}. {\itshape Phys. Rev. Lett.}
  2007, \textbf{98}, 044502.

\bibitem[\protect\citeauthoryear{{Morin}}{2005}]{Mor05}
{Morin}, V., {Instabilit\'es et bifurcations associ\'es \`a la mod\'elisation
  de la g\'eodynamo}. PhD thesis, Universit\'e Denis Diderot- Paris 7 2005.

\bibitem[\protect\citeauthoryear{{Nataf} {\itshape{et~al.}}}{2006}]{Nat06}
{Nataf}, H.C., {Alboussi\`ere}, T., {Brito}, D., {Cardin}, P., {Gagni\`ere},
  N., {Jault}, D., {Masson}, J.P. and {Schmitt}, D., Experimental study of
  super-rotation in a magnetostrophic spherical Couette flow. {\itshape
  Geophys. Astrophys. Fluid Dyn.} 2006, \textbf{100}, 281--298.

\bibitem[\protect\citeauthoryear{{Nataf} {\itshape{et~al.}}}{2008}]{Nat08}
{Nataf}, H.C., {Alboussi{\`e}re}, T., {Brito}, D., {Cardin}, P.,
  {Gagni{\`e}re}, N., {Jault}, D. and {Schmitt}, D., {Rapidly rotating
  spherical Couette flow in a dipolar magnetic field: An experimental study of
  the mean axisymmetric flow}. {\itshape Phys. Earth and Planet. Int.} 2008,
  \textbf{170}, 60--72.

\bibitem[\protect\citeauthoryear{{Pais} and {Jault}}{2008}]{Pai08}
{Pais}, M.A. and {Jault}, D., {Quasi-geostrophic flows responsible for the
  secular variation of the Earth's magnetic field}. {\itshape Geophy. J. Int.}
  2008, \textbf{173}, 421--443.

\bibitem[\protect\citeauthoryear{{Peffley} {\itshape{et~al.}}}{2000}]{Pef00}
{Peffley}, N.L., {Cawthorne}, A.B. and {Lathrop}, D.P., {Toward a
  self-generating magnetic dynamo: The role of turbulence}. {\itshape Phys.
  Rev. E} 2000, \textbf{61}, 5287--5294.

\bibitem[\protect\citeauthoryear{{Ponty} {\itshape{et~al.}}}{2007}]{Pon07}
{Ponty}, Y., {Mininni}, P.D., {Pinton}, J.F., {Politano}, H. and {Pouquet}, A.,
  {Dynamo action at low magnetic Prandtl numbers: mean flow versus fully
  turbulent motions}. {\itshape New J. Phys.} 2007, \textbf{9}, 296.

\bibitem[\protect\citeauthoryear{{Schaeffer} and
  {Cardin}}{2005{\natexlab{a}}}]{Sch05}
{Schaeffer}, N. and {Cardin}, P., {Quasigeostrophic model of the instabilities
  of the Stewartson layer in flat and depth-varying containers}. {\itshape
  Phys. Fluids} 2005{\natexlab{a}}, \textbf{17}, 104111.

\bibitem[\protect\citeauthoryear{{Schaeffer} and
  {Cardin}}{2005{\natexlab{b}}}]{Sch05b}
{Schaeffer}, N. and {Cardin}, P., {Rossby-wave turbulence in a rapidly rotating
  sphere}. {\itshape Nonlinear Process. Geophys.} 2005{\natexlab{b}},
  \textbf{12}, 947--953.

\bibitem[\protect\citeauthoryear{Schaeffer and Cardin}{2006}]{Sch06}
Schaeffer, N. and Cardin, P., {Quasi-geostrophic kinematic dynamos at low
  magnetic Prandtl number}. {\itshape Earth Planet. Sci. Lett.} 2006,
  \textbf{245}, 595--604.

\bibitem[\protect\citeauthoryear{{Sisan} {\itshape{et~al.}}}{2004}]{Sis04}
{Sisan}, D.R., {Mujica}, N., {Tillotson}, W.A., {Huang}, Y.M., {Dorland}, W.,
  {Hassam}, A.B., {Antonsen}, T.M. and {Lathrop}, D.P., {Experimental
  Observation and Characterization of the Magnetorotational Instability}.
  {\itshape Phys. Rev. Lett.} 2004, \textbf{93}, 114502.

\bibitem[\protect\citeauthoryear{Stewartson}{1966}]{Ste66}
Stewartson, K., On almost rigid rotations. Part 2. {\itshape J. Fluid Mech.}
  1966, \textbf{26}, 131--144.

\bibitem[\protect\citeauthoryear{{Stieglitz} and {M{\"u}ller}}{2001}]{Sti01}
{Stieglitz}, R. and {M{\"u}ller}, U., {Experimental demonstration of a
  homogeneous two-scale dynamo}. {\itshape Phys. Fluids} 2001, \textbf{13},
  561--564.

\bibitem[\protect\citeauthoryear{{Willis} and {Barenghi}}{2002}]{Wil02}
{Willis}, A.P. and {Barenghi}, C.F., {A Taylor-Couette dynamo}. {\itshape
  Astron. Astrophys.} 2002, \textbf{393}, 339--343.

\end{thebibliography}

\end{document}